\documentclass[JHEP,12pt]{article}

\usepackage{jheppub}
\usepackage{amssymb,amsmath,amsthm}
\usepackage[shortlabels]{enumitem}
\usepackage{braket}
\usepackage{bbold}
\usepackage{comment} 
\usepackage{subcaption}
\usepackage{graphicx, xcolor, varwidth}
\usepackage{color}
\usepackage{dsfont}

\author[1,2]{Geoff Penington}
\author[1]{Elisa Tabor}

\affiliation[1]{Leinweber Institute for Theoretical Physics and Department of Physics,\\
University of California, Berkeley, California 94720, U.S.A.}
\affiliation[2]{Leinweber Institute for Theoretical Physics, Stanford, CA 94305, U.S.A. } 
\emailAdd{geoffp@stanford.edu}
\emailAdd{etabor@berkeley.edu}

\title{The algebraic structure of gravitational scrambling}

\abstract{We introduce a new algebraic framework to describe gravitational scrambling, including the semiclassical limit of any out-of-time-order correlation function that is built out of operator insertions separated by approximately the scrambling time. In two dimensions, the scrambling algebra, which we call a modular-twisted product, is defined in terms of two copies of the Leutheusser-Liu half-sided modular inclusion of von Neumann algebras; these describe early- and late-time operators respectively. In limits where the separation between insertions is taken to be either significantly greater or smaller than the scrambling time, the modular-twisted product reduces, respectively, to free- and tensor-product algebras that were previously studied in \cite{Chandrasekaran:2022eqq}. In a sense, the modular-twisted product interpolates between these two products. Including the Hamiltonian in the scrambling algebra leads to a Type II$_\infty$ von Neumann algebra with finite renormalized entropies that interpolate between single-QES and multi-QES phases. We also describe how to generalize the modular-twisted product algebra to higher dimensions, including spatially localized boundary excitations.}

\makeatletter
\gdef\@fpheader{\mbox{}}
\makeatother

\def\H{\mathcal{H}}
\def\O{\mathcal{O}}
\def\L{\mathcal{L}}
\def\B{\mathcal{B}}

\def\A{\mathcal{A}}

\def\r{\mathcal{R}}

\def\R{\mathbb{R}}

\def\ii{\mathrm{i}}

\def\<{\langle}
\def\>{\rangle}
\def\p{\partial}
\def\tr{\rm{tr}}

\def\tfd{|\rm{TFD}\>}

\begin{document}
\maketitle

\section{Introduction}
In recent years, significant progress in our understanding of quantum gravity has come from studying the algebraic structure of gravitational observables in various controlled semiclassical limits. Most notably, we have discovered that null translations of black hole horizon can be understood as a particular half-sided modular translation of large $N$ boundary algebras \cite{Leutheusser:2021frk, Leutheusser:2021qhd} and that the Bekenstein-Hawking entropy appears as a contribution to the entropy of a crossed-product algebra constructed by adding the ADM Hamiltonian to an algebra of quantum field theory observables at asymptotic infinity \cite{Witten:2021unn, Chandrasekaran:2022eqq}.

One important feature of black hole physics is gravitational scrambling \cite{Dray:1984ha, tHooft:1987vrq, tHooft:1990fkf, Kabat:1992tb, Schoutens:1993hu, Kiem:1995iy, Cornalba:2006xk}, where out-of-time order correlators (OTOCs) rapidly decay to zero when separated by more than the scrambling time $t_{\rm scr} = (\beta/2\pi)\log S_{BH}$ \cite{Shenker:2013pqa, Shenker:2013yza, Shenker:2014cwa}. In the semiclassical, or large $N$, limit, the Bekenstein-Hawking entropy $S_{BH}$ diverges, as does the scrambling time in units where the inverse temperature $\beta$ is fixed as $N \to \infty$. As a result, scrambling cannot be seen by the large $N$ algebras defined in \cite{Leutheusser:2021frk, Leutheusser:2021qhd, Witten:2021unn}, which are constructed by taking the large $N$ limit of correlation functions of single-trace operators, with the time separations between operator insertions and the inverse temperature $\beta$ both held fixed as $N \to \infty$.

However, as was pointed out in \cite{Chandrasekaran:2022eqq}, we can construct other large $N$ algebras that are sensitive to scrambling. In the simplest case, we consider correlation functions containing both operators inserted at arbitrary finite time $t$ (with $t/\beta$ fixed as $N\to\infty$) and also operators inserted at $T(N) + t'$ with $t'$ again arbitrary but finite, while $T(N)$ is a fixed function that diverges as $N \to \infty$.\footnote{For most of this paper, we primarily focus on theories with $1+1$ bulk dimensions, where boundary operators depend only on a location in time and not in space. We discuss higher dimensions and spatial localization in Section \ref{sec:local}.} The structure of the resulting algebra depends sharply on whether $T(N)$ diverges slower or faster than the scrambling time $t_{\rm scr}$. In the former case, commutators between early- and late-time operators vanish in the $N \to \infty$ limit and the algebra becomes a tensor product. When the time separation is much larger than the scrambling time, however, all out-of-time-order correlators vanish unless the operators in question have nonzero one-point functions. Rather than the early- and late-time operators commuting, the large $N$ algebra becomes a free product of early- and late-time subalgebras.

An unanswered question in \cite{Chandrasekaran:2022eqq} was what happens when $T(N)$ is exactly equal to the scrambling time $t_{\rm scr}$. This algebra should be much richer than either the free or tensor product because it has to capture the full transition from exponentially small commutators $[a(t), b(T + t')]$ when $t' \ll t$ to exponentially small OTOCs $\braket{a(t) b(T+ t') a(t) b(T + t')}$ when $t' \gg t$. In other words, it has to capture the full physics of semiclassical gravitational scrambling in a single mathematical structure.

The purpose of the present paper is to construct exactly that structure, which (in two dimensions) we call a modular-twisted product. We define this product starting from two half-sided modular inclusions of von Neumann algebras $\widetilde\A \subseteq \A$ and $\widetilde\B\subseteq \B$. Physically, $\A$ describes operators $a(t)$ at finite time $t$, with $\widetilde\A$ the restriction to $t > 0$, while $\B$ is built out of operators $b(t_{\rm scr} + t')$ inserted at a finite time separation $t'$ from the scrambling time $t_{\rm scr}$, with $\widetilde\B$ the restriction to $t' < 0$. At intermediate times $0 \ll t \ll t_{\rm scr}$, the modes that can be excited by $\A$ are all left-moving, while the modes excited by $\B$ are all right-moving. The modular translation generator $P_A$ associated to the inclusion $\widetilde\A \subseteq \A$ translates those left-moving modes along the black hole horizon, while the generator $P_B$ associated to $\widetilde\B\subseteq \B$ translates the right-moving modes along the white hole horizon.

In the semiclassical limit of interest, the two-dimensional gravitational $S$-matrix describing the scattering of the left- and right-moving modes consists of an eikonal phase proportional to the product $P_AP_B$ of these null momenta. If a late-time boundary operator $b_R \in \B_R$ is conjugated by this gravitational $S$-matrix, it becomes an operator acting on early-time right-moving modes deep in the black hole interior. These modes commute with the early-time boundary algebra $\A_R$. In other words, we have
\begin{align}\label{eq:conjtocommute}
    [\A_R, e^{+ \ii P_A  P_B}\B_R e^{- \ii P_A  P_B}] = 0\,.
\end{align}
The modular-twisted product is, somewhat heuristically, the algebra generated by $\A_R$ and $\B_R$ subject to the condition \eqref{eq:conjtocommute}.

The description of gravitational scrambling using a modular-twisted product algebra gives an explicit connection between the saturation of the chaos bound \cite{Maldacena:2015waa} in gravitational scrambling and the rigid algebraic structure associated to half-sided modular translations. The mathematical similarity between these structures was noted already in \cite{Ceyhan:2018zfg} and significant effort has gone into trying to understand that relationship better; for a probably incomplete list of references, see e.g. \cite{Czech:2019vih, DeBoer:2019kdj, Chandrasekaran:2021tkb, Ouseph:2023juq, deBoer:2025rxx}. However, to the best of our knowledge, our work marks the first time that the Lyapunov growth of commutation relations in gravitational OTOCs has been directly related to the growth of a half-sided modular translation generator under modular flow. An interesting open question -- that goes beyond the scope of this paper -- is whether this twisted-product structure is specific to gravitational scrambling or if, perhaps, it arises in any large $N$ theory that saturates the chaos bound.

The layout of the paper is as follows. In Section \ref{sec:main}, we give a precise definition of the modular-twisted product and prove its basic properties. In Section \ref{sec:limits}, we show that in appropriate limits the modular-twisted product reduces to the tensor and free products discussed in \cite{Chandrasekaran:2022eqq}. In Section \ref{sec:typeii}, we show that adding the gravitational Hamiltonian to the modular-twisted product algebra leads to a Type II von Neumann algebra with entropies that interpolate between single-QES and multi-QES phases. Finally, in Section \ref{sec:local}, we explain how to generalize the modular-twisted product to higher-dimensional gravitational scrambling, including localized excitations.

\section{An algebra for scrambling}\label{sec:main}
\subsection{Background and review}\label{sec:review}
Following \cite{Leutheusser:2021frk, Leutheusser:2021qhd}, we first construct a von Neumann algebra describing the large $N$ limit of correlation functions for the thermofield double state $\tfd$.\footnote{See \cite{Witten:2018zxz, Sorce:2023fdx} for reviews of von Neumann algebras aimed at physicists.} Given an appropriately normalized, single-trace operator $a_0(t)$, we can define the expectation-subtracted operator
\begin{align}\label{eq:singletrace}
    a(t) = a_0(t) - \braket{a_0(t)}_{\beta}
\end{align}
where $\braket{a_0(t)}_{\beta}$ is a thermal expectation value. In the large $N$ limit, thermal two-point functions of operators of the form \eqref{eq:singletrace} are finite, while higher-point connected correlation functions vanish, as, by construction, do one-point functions. 

There is a finite subset of single-trace operators $h$ that generate symmetries of the theory that are preserved by the thermal ensemble. These operators are central at large $N$, in the sense that 
\begin{align}\label{eq:central}
    \lim_{N\to\infty}\braket{h a(t)}_{\beta} = \lim_{N\to \infty}\braket{a(t) h}_\beta
\end{align} 
for all single-trace operators $a(t)$. The most prominent example of such an operator is the rescaled Hamiltonian $h = (H - \braket{H}_\beta)/N$. For the moment, we will ignore these conserved charges and we will exclude them from the set of single-trace operators in all future discussion unless stated otherwise.

We can use the $N \to \infty$ limit of single-trace correlation functions to define a large $N$ GNS Hilbert space $\H_A$. States in this Hilbert space are defined formally in terms of products of single-trace operators acting on a GNS vacuum $\ket{\Psi_A}$ that describe the physics, in the large $N$ limit, of those excitations acting on $\tfd$. Thanks to the AdS/CFT correspondence, they also describe the physics of QFT excitations (including free graviton excitations) on a fixed two-sided black hole background. The single-trace operators \eqref{eq:singletrace} act naturally on $\H_A$; from a bulk perspective they act as local QFT operators at the right boundary. Their double commutant is a Type III$_1$ von Neumann factor $\A$ that includes all QFT observables in the right exterior. 

The commutant algebra $\A'$ can be identified with the algebra of single-trace operators acting on $\tfd$ at the left boundary or, in the bulk, with left-exterior QFT observables. Because thermal correlation functions satisfy the KMS condition, the modular Hamiltonian $K = -\log \Delta_A$ for the GNS vacuum state $\ket{\Psi_A}$ generates forwards time evolution of $\A$ and backwards time evolution of the left boundary algebra $\A'$. In the bulk, it generates global boosts of the black hole background.

The algebra $\A$ contains a proper subalgebra $\widetilde\A$ generated by single-trace operators acting at time $t > 0$. At finite $N$, such an algebra would include the Hamiltonian $H$ and hence, by conjugating by $\exp(\ii H \Delta t)$ would include all single-trace operators acting at time $t < 0$. However, in the large $N$ limit, the rescaled Hamiltonian $h = \left(H - \braket{H}\right)/N$ is central: it was therefore excluded from the algebra $\A$ and $\widetilde\A$ can be a proper subalgebra.\footnote{Even if it was included, the centrality of $h$ means that the action of $\exp\left(\ii h t \right)$ on $\A$ by conjugation is trivial, so that it could not be used to generate time evolution.} 

From a bulk perspective, the subalgebra $\widetilde\A$ describes the causal wedge of the $t>0$ boundary, i.e. right exterior fields that do not cross the horizon before infalling time $v = 0$. Its modular Hamiltonian $\widetilde K = - \log \widetilde\Delta_A$ generates boosts of the black hole horizon around the $v=0$ cut. Because the modular flow generated by $K$ preserves the subalgebra $\widetilde\A$ for positive times $t \geq 0$, we say that $\widetilde\A\subseteq \A$ is a (positive) half-sided modular inclusion. A consequence of this property is that the modular translation operator $2 \pi P_A =  K_A - \widetilde K_A$ is positive and satisfies $[K_A, P_A] = -2 \pi \ii P_A$. In the bulk, $P_A$ generates null translations of the horizon. Note that the actions of both $P_A$ and $\widetilde K_A$ are only local on the black hole horizon, where they respectively generate translation and boost symmetries. Away from the horizon (and in particular at the boundary), the background spacetime breaks those symmetries and the action of $P_A$ and $\widetilde K_A$ is nonlocal and quite complicated.

A useful fact is that the state $\ket{\Psi_A}$ is standard for the inclusion $\widetilde\A \subseteq \A$, meaning that $\ket{\Psi_A}$ is cyclic not only for  $\widetilde\A$ and $\A$, but also for $\widetilde\A' \cap \A$. This follows from standard properties of the bulk free field theory, where $\ket{\Psi_0}$ is identified with the Hartle-Hawking vacuum and $\widetilde\A' \cap \A$ with the algebra associated to a finite interval of the black hole horizon. A slightly stronger statement is that the $*$-algebra
\begin{align}\label{eq:densesubalgebra}
   \widehat\A = \bigcup_{t > 0} \Delta_A^{-\ii t}\left(\widetilde\A' \cap \A\right)\Delta_A^{\ii t}
\end{align}
is dense in $\A$ with respect to the strong operator topology (s.o.t.). In the bulk, the algebra $\Delta_A^{-\ii t}\left(\widetilde\A' \cap \A\right)\Delta_A^{\ii t}$ consists of operators localized on the black hole horizon prior to infalling time $v = t$. The statement that $\widehat\A$ is s.o.t. dense in $\A$ captures the fact that right-exterior operators can be approximated, to arbitrary precision, by operators localized on the black hole horizon within a finite range of affine times. Again, this is a standard property of free quantum field theory applied to an AdS-black hole background.\footnote{Note that it was crucial here that we excluded the central operators satisfying \eqref{eq:central} from the algebra $\A$ and hence that the bulk QFT dual to $\H_A$ does not include nontrivial superselection sectors.}

Our primary object of interest is not $\H_A$. Instead, it is a larger GNS Hilbert space, similar to those first introduced in \cite{Chandrasekaran:2022eqq}. This Hilbert space is built from correlation functions that include not only finite-time single-trace operators $a(t)$ but also single-trace operators $b(t' + T(N))$, where $t'$ is arbitrary but finite, while $T(N)$ is a fixed function of $N$ that diverges (in units with $\beta$ fixed) as $N \to \infty$. Because black holes thermalize, large $N$ two-point functions $\braket{a(t) b(T + t')}_\beta$ go to zero as $T \to \infty$. In fact, more generally, so long as $\beta \ll T \ll t_{\rm scr}$, all correlation functions factorise as
\begin{align}
    \lim_{N \to \infty} \braket{a_1(t_1) b_1(T+ t_1') a_2(t_2) \dots}_\beta = \braket{a_1(t_1)a_2( t_2) \dots}_\beta \braket{b_1( t_1')b_2(t_2')\dots}_\beta.
\end{align}
As a result, the GNS Hilbert space factorises as $\H \cong \H_A \otimes \H_B$ with $\H_B$ and $\H_A$ isomorphic but with the late-time algebra $\B$ acting on $\H_B$ while the early-time algebra $\A$ acts on $\H_A$. The full right-boundary algebra is the tensor product $\A \otimes \B$. The thermofield double $\tfd$ is identified with the GNS vacuum $\ket{\Psi} = \ket{\Psi_A} \ket{\Psi_B}$. 

At times $0 \ll t \ll T(N)$, all the modes in $\H_A$ (which includes both the Hawking modes that can be excited directly by $\A$ and their interior partners) are left-moving, while the modes in $\H_B$ are right-moving; this provides convenient terminology to distinguish the two (that we will also adopt for $T(N) \approx t_{\rm scr}$), even though it is only technically true within a limited range of boundary times.

On the other hand, when $T(N) \gg t_{\rm scr}$, all out-of-time-order correlators
\begin{align}\label{eq:OTOCvanish}
    \braket{a_1 b_1 a_2 \dots}_\beta \to 0
\end{align}
vanish if $\braket{a_1}_\beta = \braket{b_1}_\beta=  \braket{a_2}_\beta \dots = 0$.\footnote{The leftmost and rightmost operators in \eqref{eq:OTOCvanish} may each be either early- or late-time operators.} This leads to a much larger GNS Hilbert space that can be written as
\begin{align}
    \H \cong \ket{\Psi} \oplus \H_A^* \oplus \H_B^* \oplus \left(\H_A^* \otimes \H_B^*\right) \oplus \left(\H_B^* \otimes \H_A^*\right) \oplus \left(\H_A^* \otimes \H_B^* \otimes \H_A^*\right) \oplus\dots
\end{align}
where $\ket{\Psi}$ is the GNS vacuum (identified with the thermofield double state), $\H_A^* \cong \H_A \ominus \ket{\Psi_A}$ (and  $\H_B^* \cong \H_B \ominus \ket{\Psi_B}$) describes all excited states in $\H_A$ (and $\H_B$), and the infinite sum is over all alternating tensor products of 
$\H_A^*$ and $\H_B^*$. The right boundary algebra is the so-called free product of $\A$ and $\B$, which is the double commutant of $\A \cup \B$ on the GNS Hilbert space and includes arbitrary alternating products of $\A$ and $\B$. In the bulk, states in a subspace such as $\left(\H_A^* \otimes \H_B^* \otimes \H_A^*\right)$ describe a sequence of high-energy shocks (in this case three) supporting a long wormhole. Each new shock creates a large backreaction that entirely hides the previous shock behind the horizon. Right boundary operators either act on the rightmost shock (if that shock is close to them in time) or they create a new shock if the rightmost shock is at late times and the operator acts at early times (or vice versa).

\subsection{The modular-twisted product}
We are interested in the case where early- and late-time excitations are separated by exactly a scrambling time (up to an irrelevant finite correction), i.e., with 
\begin{equation}\label{eq:TN=tscr}
    T(N) = \frac{\beta}{2\pi}\log N^2\,+ O(1)\,.
\end{equation}
For convenience, we will henceforth work in units where the black hole inverse temperature $\beta$ is equal to one. As discussed, for the moment we will also assume two bulk spacetime dimensions in order to avoid complications that arise from the spatial localization of excitations.

 Unlike in the limits discussed above, when \eqref{eq:TN=tscr} holds, the interaction between the left- and right-moving excitations is sufficient to create a nontrivial backreaction, but is not strong enough to guarantee that any left-moving excitation will hide all prior right-moving excitations far behind the black hole horizon (or vice versa). Instead, the interaction between the excitations leads to an eikonal phase
$\delta \propto G_N s$, where $s$ is the center of mass energy.\footnote{This phase can be derived by summing over crossed-ladder graviton-exchange diagrams, which are the only ones to survive the limit of interest. See \cite{Shenker:2014cwa} for a simple explanation and further references.}

\begin{figure}[t]
    \centering
    \includegraphics[width=16cm]{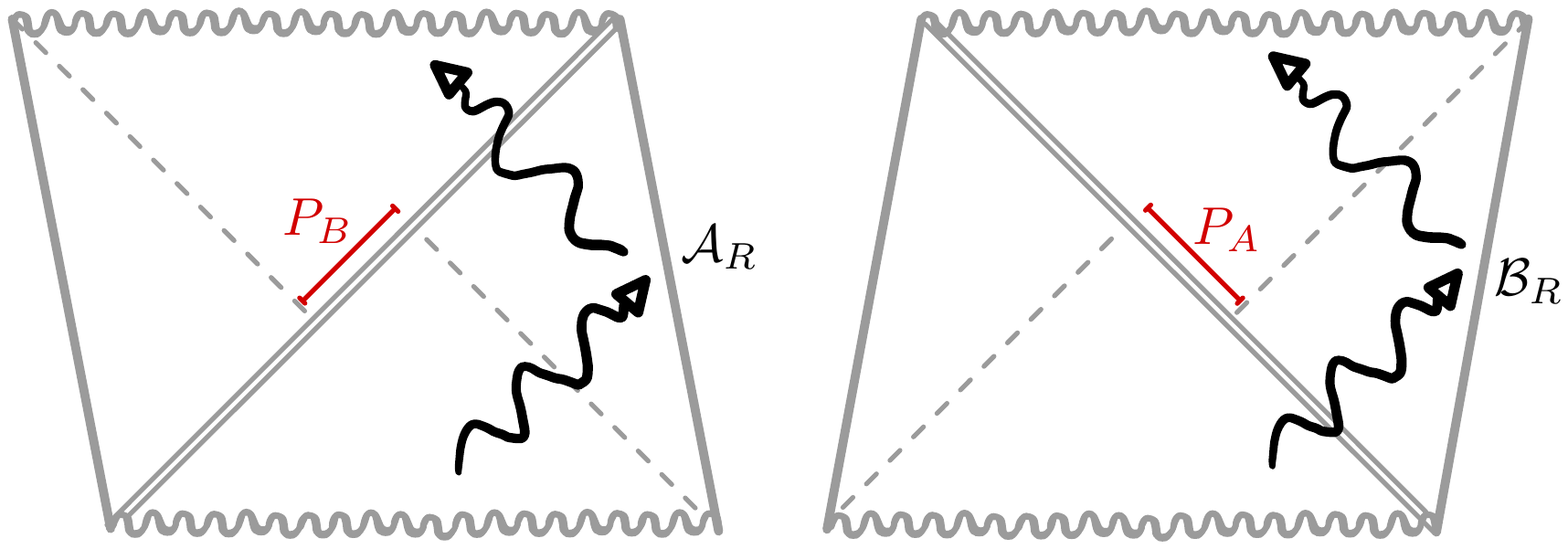}
    \vspace{-5.4cm}
    \caption{In a boost frame in which $\H_A$ modes have low energies (left), excitations of $\H_B$ modes create a high-energy right-moving shock that creates a null shift of size $P_B$ on the black hole horizon. On the other hand, in a boost frame where $\H_B$ modes have low energies (right), excitations of $\H_A$ modes create a high-energy left-moving shock of size $P_A$ on the white hole horizon. Both perspectives explain the same eikonal phase $\hat \delta \propto P_AP_B$ that results from the interaction of the two shocks. }
    \label{fig:PAPB}
\end{figure}

Up to corrections that are subleading as $T \to \infty$, we have $s \propto e^{2\pi T} P_A P_B$ where $P_A$ is the null energy of the left-moving modes crossing the black hole horizon, while $P_B$ is the null energy of the right-moving modes on the white hole horizon. When $T$ is given by \eqref{eq:TN=tscr}, the factor of $e^{2\pi T}$ cancels the factor of $G_N \sim N^{-2}$ so that the eikonal phase remains finite as $N \to \infty$. For convenience, we can choose the finite piece of $T$ to absorb any remaining finite prefactor so that the eikonal phase becomes
\begin{align}
    \hat\delta = P_A P_B\,.
\end{align}
From a semiclassical perspective, this phase can be understood as coming from the translation of the left-moving modes due to backreaction from the right-moving modes or vice versa; see Figure \ref{fig:PAPB}.

The operator $P_A$ can be understood from a boundary perspective by recognising that it is precisely the modular translation operator $2\pi P_A =  K_A - \widetilde K_A$ associated to the (positive) half-sided modular inclusion $\widetilde\A \subseteq \A$ of operators at time $t > 0$ described in Section \ref{sec:review}. The operator $P_B$ is a time reversal of $P_A$: it is the modular translation $2\pi P_B = K_B  -\widetilde K_B  $ associated to the half-sided modular inclusion $\widetilde \B \subseteq \B$ of late-time operators with $t' < 0$. This is a negative half-sided modular inclusion, meaning that $\Delta_B^{-\ii t}\widetilde \B \Delta_B^{\ii t} \subseteq \B$ for $t < 0$. As a result, we have $[K_B, P_B] = 2 \pi \ii P_B$.

\begin{figure}[t]
    \centering
    \includegraphics[width=12cm]{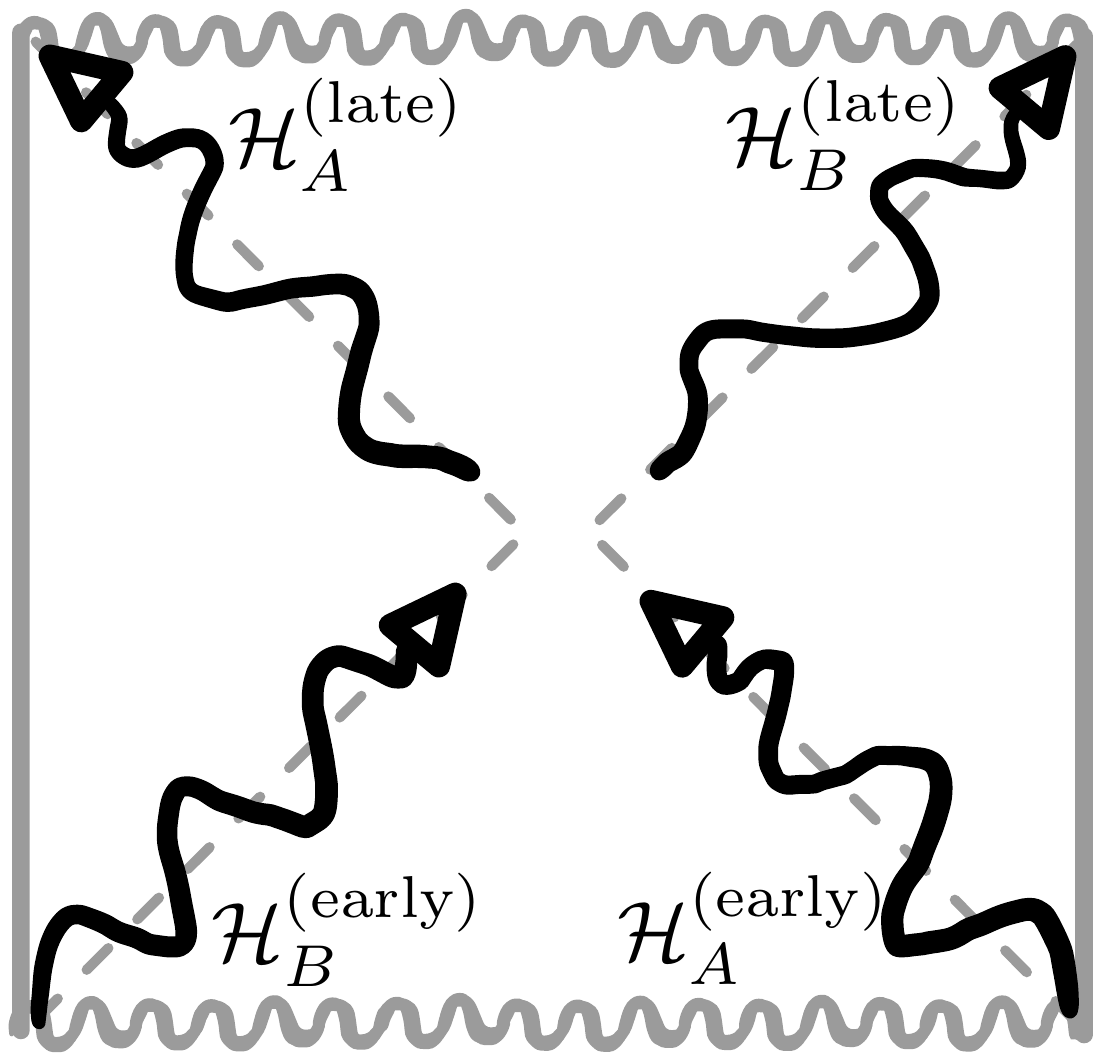}
    \vspace{-.8cm}
    \caption{The bulk Hilbert space of early-time excitations, related to the Hilbert space of late-time excitations by the unitary $\exp(\ii P_A P_B)$.}
    \label{fig:modes}
\end{figure}

At early times, excitations of left- and right-moving modes are spatially separated by a parametrically large distance. The bulk Hilbert space can therefore be written as a tensor product
\begin{align}
    \H \cong \H_A^{\rm (early)} \otimes \H_B^{\rm (early)},
\end{align}
where the early-time left-moving modes $\H_A^{\rm (early)}$ are acted on by $\A_R$ and are located near the right white hole horizon while the early-time right-moving modes $\H_B^{\rm (early)}$ are near the left white hole horizon. We can construct a similar decomposition at late times, writing
\begin{align}
    \H \cong \H_A^{\rm (late)} \otimes \H_B^{\rm (late)},
\end{align}
where the late-time right-moving modes $\H_B^{\rm late}$, which are acted on by $\B_R$, are located near the right black hole horizon, while the late-time left-moving modes $\H_A^{\rm late}$ are located near the left white hole horizon; see Figure \ref{fig:modes}.

However, the nontrivial scattering that occurs between the left- and right-moving modes near the bifurcation surface means that these two decompositions are not the same. Instead they are related by precisely the unitary $\exp(\ii P_A P_B)$ that, as we explained above, describes elastic gravitational eikonal scattering. It is convenient to pick one canonical decomposition of $\H$ to relate all the others to: rather than break time reflection symmetry by picking either the early- or late-time decomposition, we will split the difference, writing
\begin{align}\label{eq:differentdecomps}
    \H \cong \H_A \otimes \H_B = e^{\ii P_A P_B/2} \H_A^{\rm (early)} \otimes \H_B^{\rm (early)} = e^{-\ii P_AP_B/2} \H_A^{\rm (late)} \otimes \H_B^{\rm (late)}.
\end{align}
The thermofield double state $\tfd$ is identified with the GNS vacuum state $\ket{\Psi} = \ket{\Psi_A}\ket{\Psi_B}$.\footnote{Note that, since $P_A \ket{\Psi_A} = P_B\ket{\Psi_B} = 0$, the relation $\ket{\Psi} = \ket{\Psi_A}\ket{\Psi_B}$ holds in all three decompositions given in \eqref{eq:differentdecomps}. In fact, the three decompositions in \eqref{eq:differentdecomps} are all equivalent whenever we project either set of modes into the vacuum state (i.e. after applying either $\ket{\Psi_A}\!\bra{\Psi_A}$ or $\ket{\Psi_B}\!\bra{\Psi_B}$).}

The right boundary algebra $\r$ is generated by operators $a_R \in \A_{R}$ acting on $\H_A^{\rm (early)}$ together with operators  $b_R \in \B_{R}$ acting on $\H_B^{\rm (late)}$. In other words, it is generated by operators of the form
\begin{equation} \label{eq:boostedelements}
\begin{split}
    a_R &= e^{\ii P_A P_B/2} \,a\,e^{-\ii P_A P_B/2} \,,\\
    b_R &= e^{-\ii P_A P_B/2}\,b\,e^{\ii P_A P_B/2} \,, 
\end{split}
\end{equation} 
where $a \in \A$ and $b \in \B$ act on $\H_A$ and $\H_B$ respectively in the decomposition \eqref{eq:differentdecomps}. As a result, even though the Hilbert space is a tensor product, the subalgebras $\A_R = e^{\ii P_A P_B/2} \,\A\,e^{-\ii P_A P_B/2}$ and $\B_R = e^{-\ii P_A P_B/2} \,\B\,e^{\ii P_A P_B/2}$ do not commute. This reflects the gravitational interaction between the two sets of modes, or equivalently scrambling of the boundary degrees of freedom. 

Instead of the usual tensor product condition $[\A,\B] = 0$, the defining property of the algebra $\r = (\A_R\vee \B_R)''$ is the condition
\begin{align}
    [\A_R, e^{\ii P_A P_B}\B_Re^{-\ii P_A P_B}] = 0\,.
\end{align}
We therefore call the algebra $\r$ a ``modular-twisted product'' of $\A_R$ and $\B_R$.

\begin{figure}[t]
    \centering
    \includegraphics[width=12cm]{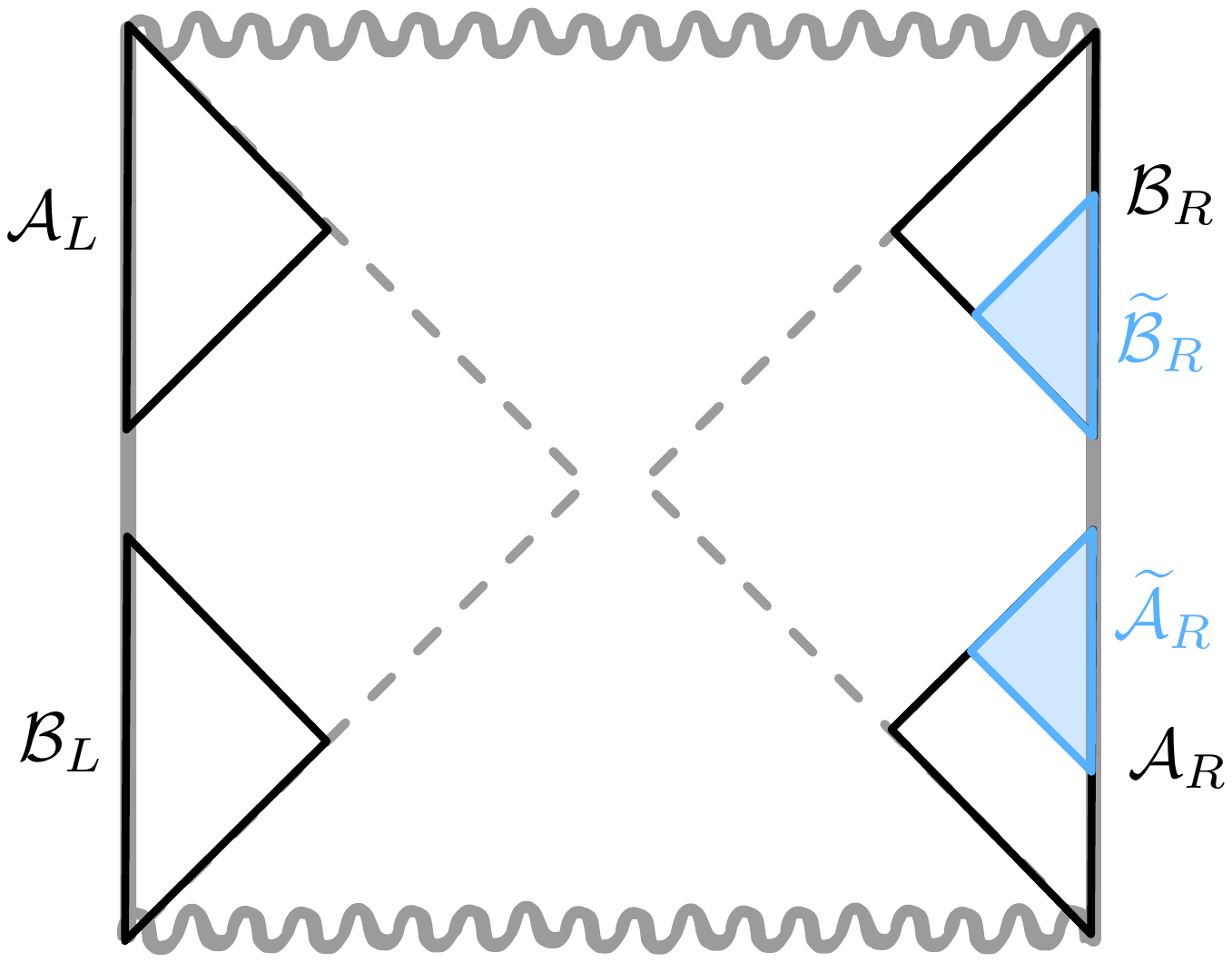}
    \vspace{-.6cm}
    \caption{The right-boundary algebra $\r$ is generated by the early-time boundary algebra $\A_R$ together with the late-time boundary algebra $\B_R$. Their respective subalgebras $\widetilde\A_R$ and $\widetilde\B_R$ define half-sided modular translations that are generated respectively by $P_A$ and $P_B$ and that help define the modular-twisted product. The commutant algebra $\L$ is generated by the early-time left-boundary algebra $\B_L$ together with the late-time left-boundary algebra $\A_L$.}
    \label{fig:subalgebras}
\end{figure}

The rest of this section will be devoted to studying properties of the algebra $\r$. In particular, we will argue that it is a Type III$_1$ von Neumann factor, with a commutant $\L = (\A_R\vee \B_R)' = (\A_L\vee \B_L)''$ that is generated by the subalgebras $\A_L = e^{-\ii P_A P_B/2} \,\A'\,e^{\ii P_A P_B/2}$ and $\B_L = e^{\ii P_A P_B/2} \,\B'\,e^{-\ii P_A P_B/2}$. Physically, $\A_L$ can be interpreted as operators acting on the left boundary at late times (and hence on left-moving modes), while $\B_L$ describes operators acting at the left boundary at early times (and hence on right-moving modes); see Figure \ref{fig:subalgebras}. 

\subsection{Properties}
\subsubsection*{$\A_L$ and $\B_L$ commute with $\r$}
Given $a_L = e^{-\ii P_A P_B/2} a'\,e^{\ii P_A P_B/2} \in \A_L$ and $b_R = e^{-\ii P_A P_B/2} b\,e^{\ii P_A P_B/2} \in \B_R$, we have
\begin{align}
    [a_L, b_R] = e^{-\ii P_A P_B/2} [a', b] e^{\ii P_A P_B/2} = 0\,.
\end{align}
So $\A_L$ commutes with $\B_R$. Given $a_R = e^{\ii P_A P_B/2} a\,e^{-\ii P_A P_B/2} \in \A_R$, we have
\begin{align}
    [a_L, a_R] = \int_0^\infty dp_B \pi_B(p_B) e^{-\ii P_A p_B/2} [a', e^{\ii P_A p_B}a e^{-\ii P_A p_B}] e^{\ii P_A p_B/2} = 0\,,
\end{align}
where $dp_B \pi_B(p_B)$ is the projection-valued measure for the positive operator $P_B$. In the second equality we used the fact that $P_A$ generates a half-sided modular translation so that $e^{\ii P_A p_B}a e^{-\ii P_A p_B}$ is contained in $\A$ for any fixed $p_B \geq 0$.

Since $\A_L$ commutes with both $\A_R$ and $\B_R$, it also commutes with $\r = (\A_R \vee \B_R)''$. The argument that $\B_L$ commutes with $\r$ is identical. We conclude that the left boundary algebra $\L = (\A_L \vee \B_L)''$ commutes with $\r$. (We have not yet shown that $\L$ is the full commutant algebra $\r'$.)

\subsubsection*{$|\Psi\>$ is cyclic for $\r$}
Given $a_{R,i} = e^{\ii P_A P_B/2} a_{i} e^{-\ii P_A P_B/2}\in \A_R$ and $b_{R,i} = e^{-\ii P_A P_B/2} b_{i} e^{\ii P_A P_B/2} \in \B_R$, we have
\begin{equation}\label{eq:densesetstates}
    \sum_i a_{R,i}b_{R,i}|\Psi\> = \sum_i e^{\ii P_A P_B/2} a_{i}|\Psi_A\> b_{i}|\Psi_B\>\,.
\end{equation}
Since states of the form $a_{i}|\Psi_A\>$ are dense in $\H_A$, states of the form $b_{i}|\Psi_B\>$ are dense in $\H_B$, and $e^{\ii P_A P_B/2}$ is unitary, states of the form \eqref{eq:densesetstates} are dense in $\H$.

\subsubsection*{$|\Psi\>$ is separating for $\r$}
The vacuum $|\Psi\>$ is separating for $\r$ if  and only if it is cyclic for the commutant algebra $\r'$, which we have shown contains both $\A_L$ and $\B_L$, along with products thereof. Given $a_{L,i} = e^{-\ii P_A P_B/2} a_{i}' e^{\ii P_A P_B/2}\in \A_L$ and $b_{L,i} = e^{\ii P_A P_B/2} b_{i}' e^{-\ii P_A P_B/2} \in \B_R$, we have
\begin{equation}
    \sum_i a_{L,i}b_{L,i}|\Psi\> = \sum_i e^{-\ii P_A P_B/2} a_{i}'|\Psi_A\> b_{i}'|\Psi_B\>\,.
\end{equation}
States of this form are dense in $\H$ by an identical argument to the one above.

\subsubsection*{$P_A P_B$ commutes with a product of modular flows}
Let $\Delta_A$ and $\Delta_B$ be respectively the modular operators for $\ket{\Psi_A}$ on $\A$ and $\ket{\Psi_B}$ on $\B$ respectively. Recall that (e.g. Eq. (5.23-26) of \cite{Leutheusser:2021frk})
\begin{equation}\label{eq:DeltaPalgebra}
    [\log\Delta_A,P_A] = 2\pi\ii P_A\,,\quad [\log\Delta_B,P_B] = -2\pi\ii P_B\,.
\end{equation}
Hence
\begin{equation}
    [\log\Delta_A,P_A P_B] = 2\pi\ii P_A P_B\,,\quad [\log\Delta_B,P_A P_B] = -2\pi\ii P_A P_B
\end{equation}
and
\begin{equation}\label{eq:Deltacommutes}
    [\Delta_A\otimes\Delta_B,P_A P_B] = 0\,.
\end{equation}

\subsubsection*{$P_A P_B$ commutes with a product of modular conjugations}
Let $J_A$ and $J_B$ be the (antilinear) modular conjugation operators for $\ket{\Psi_A}$ on $ \A$ and   $\ket{\Psi_B}$ on $\B$ respectively. Following Eq. (2.10) of \cite{Ceyhan:2018zfg}, we have
\begin{equation}
    J_A e^{-\ii \lambda P_A} J_A = e^{\ii \lambda P_A}\,,\quad J_B e^{-\ii \lambda P_B} J_B = e^{\ii \lambda P_B}\,.
\end{equation}
Since $J_A$ and $J_B$ are antilinear operators (mapping $\ii$ to $-\ii$) with $J_A^2 = J_B^2 = 1$, this is equivalent to
\begin{equation}\label{eq:JPAandB}
    [J_A, P_A] = 0\,,\quad [J_B, P_B] = 0\,.
\end{equation}
The tensor product $J_A\otimes J_B$ is an antilinear operator on $\H \cong \H_A \otimes \H_B$.\footnote{The easiest way to understand this is to remember that an antilinear operator on $\H$ is a linear operator from $\H$ to the conjugate Hilbert space $\overline\H$. $J_A \otimes J_B$ is therefore a linear operator from $\H_A \otimes \H_B$ to $\overline \H \cong \overline\H_A \otimes \overline\H_B$ i.e. an antilinear operator on $\H$.} From \eqref{eq:JPAandB}, we have
\begin{equation}\label{eq:Jcommutes}
    (J_A\otimes J_B) e^{\ii P_A P_B/2}(J_A\otimes J_B) = e^{-\ii P_A P_B/2}\,.
\end{equation}

\subsubsection*{The Tomita operator $S_\Psi$ extends $S_A \otimes S_B$}
Since $\ket{\Psi}$ is cyclic-separating for $\r$, we can define the Tomita operator $S_\Psi$ as the closure of
\begin{equation}
    S_\Psi r|\Psi\> = r^\dagger|\Psi\> 
\end{equation}
for $r \in \r$. This domain includes as a subspace states of the form \eqref{eq:densesetstates}. 

Now let $S_A = J_A \Delta_A^{1/2}$ and $S_B = J_B \Delta_B^{1/2}$ be the Tomita operators for $\A$ on $\ket{\Psi_A}$ and $\B$ on $\ket{\Psi_B}$ respectively. Using \eqref{eq:Deltacommutes} and \eqref{eq:Jcommutes}, we have
\begin{align}
    (S_A\otimes S_B) e^{\ii P_A P_B/2} &= (J_A\Delta_A^{1/2}\otimes J_B\Delta_B^{1/2}) e^{\ii P_A P_B/2} \nonumber\\
    &= e^{-\ii P_A P_B/2} (J_A\Delta_A^{1/2}\otimes J_B\Delta_B^{1/2}) = e^{-\ii P_A P_B/2}(S_A\otimes S_B)\label{eq:SASBPAPB} \,.
\end{align}
By construction, states of the form $\sum_i a_{i}b_{i}|\Psi\>$ form a core for $S_A \otimes S_B$, where a core is a dense set of states in the domain of $S_A \otimes S_B$ such that restricting $S_A \otimes S_B$ to the core and then taking a closure recovers $S_A \otimes S_B$. Since $e^{\ii P_A P_B/2}$ is unitary, it then follows from \eqref{eq:SASBPAPB} that states of the form \eqref{eq:densesetstates} are also a core for $S_A \otimes S_B$. Moreover, on such states, we have
\begin{align}
    (S_A\otimes S_B) \sum_i a_{R,i} b_{R,i}|\Psi\> &= (S_A\otimes S_B) e^{\ii P_A P_B/2} \sum_i a_{i} b_{i}|\Psi\> \\
    &= e^{-\ii P_A P_B/2} (S_A\otimes S_B) \sum_i a_{i}b_{i}|\Psi\> 
    \\&= e^{-\ii P_A P_B/2} \sum_i b_{i}^\dag a_{i}^\dag|\Psi\>
    \\&=\sum_i  b_{R,i}^\dagger a_{R,i}^\dagger|\Psi\>  = S_\Psi\sum_i a_{R,i} b_{R,i}|\Psi\>\,.
\end{align}
We conclude that $S_\Psi$ and $S_A \otimes S_B$ agree on a core for $S_A \otimes S_B$ and hence that $S_\Psi$ is extension of $S_A \otimes S_B$.\footnote{A closed operator $S_\Psi$ is extension of $S_A \otimes S_B$ if the domain of $S_\Psi$ contains the domain of $S_A \otimes S_B$ and the two operators agree when restricted to the latter domain.}

\subsubsection*{Tomita operator factorization}
The adjoint of the Tomita operator $S_\Psi$ is the Tomita operator $S_\Psi'$ of the commutant algebra $\r'$. In others words, $S_\Psi^\dagger$ is the closure of 
\begin{equation}\label{eq:Spsi'domain}
    S_\Psi^\dagger r'|\Psi\> = r'{}^\dag|\Psi\>
\end{equation}
for all $r' \in \r'$. Recall that $\r'$ contains the left boundary algebra $\L = (\A_L \vee \B_L)''$. The domain in \eqref{eq:Spsi'domain} therefore includes states of the form 
\begin{align}\label{eq:aLbLdensedomain}
    \sum_i a_{L,i} b_{L,i} \ket{\Psi} = e^{\ii P_A P_B/2}\sum_i  a_{i}'b_{i}'|\Psi\>
\end{align}
for $a_{L,i} = e^{\ii P_A P_B/2} a_{i}' e^{-\ii P_A P_B/2}\in \A_L$ and $b_{L,i} = e^{-\ii P_A P_B/2} b_{i}' e^{\ii P_A P_B/2} \in \B_L$.

The Tomita operators $S_A' = S_A^\dagger = \Delta_A^{1/2} J_A$ and $S_B' = S_B^\dagger = \Delta_B^{1/2} J_B$ satisfy
\begin{align}
    (S_A'\otimes S_B') e^{-\ii P_A P_B/2} =  e^{\ii P_A P_B/2}(S_A'\otimes S_B')\,.
\end{align}
So states of the form \eqref{eq:aLbLdensedomain}, like states of the form $\sum_i  a_{i}'b_{i}'|\Psi\>$, form a core for $S_A' \otimes S_B'$. We have
\begin{align}
    (S_A'\otimes S_B') \sum_i a_{L,i}b_{L,i}|\Psi\> &= (S_A'\otimes S_B') e^{\ii P_A P_B/2} \sum_i a_{i}'b_{i}'|\Psi\> \\
    &= e^{\ii P_A P_B/2} (S_A'\otimes S_B')\sum_i a_{i}'b_{i}'|\Psi\> 
    \\&= e^{-\ii P_A P_B/2} \sum_i b_{i}'{}^\dag a_{i}'{}^\dag|\Psi\>
    \\&= \sum_i b_{L,i}'{}^\dag a_{L,i}'{}^\dag|\Psi\> = S_\Psi' \sum_i a_{L,i}b_{L,i}|\Psi\> \,.
\end{align}
We conclude $S_\Psi'=S_\Psi^\dag$ extends $S_A' \otimes S_B' = S_A^\dagger \otimes S_B^\dagger$. But this is true only if $S_A \otimes S_B$ extends $S_\Psi$. We have therefore shown that in fact 
\begin{align}\label{eq:tomitafactor}
    S_\Psi = S_A \otimes S_B\,.
\end{align}

\subsubsection*{Modular operator/conjugation factorization}
It follows immediately from \eqref{eq:tomitafactor} that
\begin{align}
    \Delta_\Psi = S_\Psi^\dagger S_\Psi = S_A^\dagger S_A \otimes S_B^\dagger S_B = \Delta_A \otimes \Delta_B
\end{align}
and 
\begin{align}
    J_\Psi = \Delta_\Psi^{-1/2} S_\Psi = J_A \otimes J_B\,.
\end{align}

\subsubsection*{The left and right algebras are commutants $\r'=\L$}
For any cyclic separating state $\ket{\Psi}$ and algebra $\r$, the modular conjugation operator $J_\Psi$ defines an antilinear isomorphism from $\r$ to its commutant
\begin{align}
    \r' = J_\Psi \r J_\Psi\,.
\end{align}
However, for $a_R \in \A_R$ we have
\begin{align}
    J_\Psi a_R J_\Psi &= J_A \otimes J_B \left(e^{-\ii P_A P_B/2} \,a\,e^{\ii P_A P_B/2}\right) J_A \otimes J_B \\&\nonumber= e^{\ii P_A P_B/2} J_A a J_A e^{-\ii P_A P_B/2}\,.
\end{align}
Since $J_A \A J_A = \A'$, this means that $J_\Psi \A_R J_\Psi = \A_L$. Similarly, $J_\Psi \B_R J_\Psi = \B_L$. Finally, since $\r = (\A_R \vee \B_R)''$, we have
\begin{align}
    \r' = (J_\Psi\A_R J_\Psi \vee J_\Psi\B_RJ_\Psi)''=  (\A_L \vee \B_L)'' = \L\,.
\end{align}

\subsubsection*{The algebra $\r$ is a factor}
To prove the remaining properties of $\r$, we will make use of the fact that $\widetilde\A \subseteq \A$ and $\widetilde\B \subseteq \B$ are standard inclusions, as discussed in Section \ref{sec:review}. (As a result, our derivations, unlike those appearing previously in this section, only apply to the subclass of modular-twisted products that come from standard inclusions.)

We know from \eqref{eq:DeltaPalgebra} that any normalizable eigenstate $\ket{\Phi_A}$ of $\Delta_A$ satisfies
\begin{align}
\braket{\Phi| f(P_A) |\Phi} = \braket{\Phi|\Delta^{-is} f(P_A) \Delta^{is}|\Phi} = \braket{\Phi| f(e^{2 \pi s} P_A) |\Phi}
\end{align}
for any bounded function $f$. It follows that any such eigenstate must be in the kernel of $P_A$ i.e $\Delta_A$. But, because the inclusion $\widetilde\A \subseteq \A$ is standard, the kernel of $P_A$ consists only of $\ket{\Psi_A}$ \cite{Wiesbrock:1992mg}. We conclude that $\Delta_A$ has purely continuous spectrum except for the single eigenstate $\ket{\Psi_A}$, which it is easy to check has eigenvalue one. An identical argument applies to $\Delta_B$, which has purely continuous spectrum except for $\ket{\Psi_B}$.

Since $\Delta_\Psi = \Delta_A \otimes \Delta_B$, $\Delta_\Psi$ also contains exactly one normalizable eigenstate $\ket{\Psi} = \ket{\Psi_A} \ket{\Psi_B}$ with eigenvalue one. Suppose, however, that $\r$ contained a nontrivial central projector $r$. We would then have
\begin{align}
    \Delta_\Psi r \ket{\Psi} = S_\Psi' r^\dagger \ket{\Psi} = r \ket{\Psi} \,.
\end{align}
Since $\ket{\Psi}$ is separating, $r \ket{\Psi}$ must be linearly independent from $\ket{\Psi}$, giving our desired contradiction.

\subsubsection*{The algebra $\r$ is a Type III$_1$ factor}
The operator $P_A$ commutes with $\B_R$ and generates a half-sided modular translation for $\A_R$. We also have $P_A \ket{\Psi}  = 0$. It follows that $P_A$ generates a half-sided modular translation for the full right boundary algebra $\r$. But nontrivial half-sided modular translations can exist only for Type III$_1$ von Neumann algebras \cite{Wiesbrock:1992mg, Borchers:1993, Borchers:1998, ARAKI_2005}.

\section{Limiting behaviors}\label{sec:limits}
In this section we demonstrate that the twisted modular product algebra limits in an appropriate way to the tensor product and free product algebras that were shown in \cite{Chandrasekaran:2022eqq} to describe separations $T(N)$ far from the scrambling time.

More specifically, we show the following. For each limit, we will choose appropriate s.o.t. dense  $*$-subalgebras $\widehat{\A}_R \subseteq \A_R$ and $\widehat{\B}_R \subseteq \B_R$. Let
\begin{align}
\hat b_{R}(t) = \Delta_\Psi^{-\ii t} \hat b_R \Delta_\Psi^{\ii t} = e^{-\ii P_A P_B/2}\Delta_B^{-it}\hat b \Delta_B^{it} e^{\ii P_A P_B/2}
\end{align}
describe the time evolution of the operator $\hat b_R  = e^{-\ii P_A P_B/2} \hat b e^{\ii P_A P_B/2}$.
We can then consider all correlation functions of the form 
\begin{align}\label{eq:Ct2nddef}
    C(t) = \braket{\Psi|\hat a_{R,1} \hat b_{R,1}(t) \hat a_{R,2} \hat b_{R,2}(t) \dots |\Psi}
\end{align}
for $a_{R,i} \in \widehat{\A}_R$ and $b_{R,i} \in \widehat{\B}_R$. For convenience, we will always assume that the leftmost operator in $C(t)$ is in $\widehat{\A}_R$ and the rightmost operator is in $\widehat{\B}_R$; the other possible cases are accommodated by setting $\hat a_{R,1} = 1$, $\hat b_{R,n} = 1$  or both. When writing \eqref{eq:Ct2nddef}, we chose to evolve the operators in $\widehat{\A}_R$ forwards by time $t$, but, because of the time translation invariance of $\ket{\Psi}$, we could equivalently have evolved the operators in $\widehat{\B}_R$ backwards by $t$ or done a combination of the two. The essential point is that the separation between the $\widehat{\A}_R$ operators and the $\widehat{\B}_R$ operators is increased by $t$.

It follows from \eqref{eq:DeltaPalgebra} that
\begin{align}
C(t) = \braket{\Psi|\hat a_1 e^{-\ii \alpha P_A P_B} \hat b_1 e^{\ii \alpha P_A P_B} \hat a_2 \dots |\Psi}\,,    
\end{align}
where $\alpha = \exp(2\pi t)$, $\hat a_i =e^{-\ii P_A P_B/2} \hat a_{R,i} e^{\ii P_A P_B/2}$ and $\hat b_i =e^{\ii P_A P_B/2} \hat b_{R,i} e^{-\ii P_A P_B/2}$.

We can use the correlation functions $C(t)$ to construct a GNS Hilbert space on which the algebras $\widehat{\A}_R$ and $\widehat{\B}_R$ act, just as was done in the original Leutheusser-Liu construction. Because $(\widehat{\A}_R \cup \widehat{\B}_R)'' = (\A_R \cup \B_R)'' = \r$ (and $\r$ is cyclic for $\ket{\Psi}$) this GNS Hilbert space will be canonically isomorphic, for any finite $t$, to $\H \cong \H_A \otimes \H_B$ (with the GNS actions of $\widehat{\A}_R$ and $\widehat{\B}_R$ identified with their original actions on $\H$ except that the latter is evolved by time $t$). 

However, we can also construct a GNS Hilbert space from the limits of the correlation functions $C(t)$ as $t \to \pm \infty$. (Our results will show that those limits exist for appropriate choices of $\widehat \A_R$ and $\widehat \B_R$.) In the limit $t \to - \infty$, the separation between early-time and late-time boundary operators becomes much less than a scrambling time. We therefore expect the gravitational scattering to be suppressed so that the right boundary algebra $(\widehat{\A}_R \cup \widehat{\B}_R)''$ becomes the tensor product algebra found for $T(N) \ll t_{\rm scr}$. On the other hand, as $t \to +\infty$, the separation becomes much larger than the scrambling time and we expect to recover the free product algebra. Both expectations will turn out to be correct.

\subsection{Tensor product limit}
We first consider the limit of $t\to - \infty$. This corresponds to the limit $\alpha \to 0$ in \eqref{eq:Ct2nddef}. In this limit, there is no need to restrict to s.o.t. dense (proper)  subalgebras; we are free to choose $\widehat\A_R = \A_R$ and $\widehat \B_R = \B_R$. 

It is easy to see that
\begin{align}\label{eq:Cttensor}
    \lim_{t \to -\infty} C(t) = \braket{\Psi|\hat a_1 \hat b_1 \hat a_2 \hat b_2\dots |\Psi} = \braket{\Psi_A|\hat a_1  \hat a_2 \dots |\Psi_A}\braket{\Psi_B|\hat b_1  \hat b_2\dots |\Psi_B} \,.
\end{align}
The convergence is guaranteed since $\exp(\ii \alpha P_A P_B) \to_{\rm s.o.t.} 1$ as $\alpha \to 0$. But the right-hand side of \eqref{eq:Cttensor} is simply the expectation value of $\hat a_1 \dots \hat a_n \otimes \hat b_1 \dots \hat b_n$ in the state $\ket{\Psi_A} \ket{\Psi_B}$. As a result, the GNS Hilbert space can still be identified with $\H_A \otimes \H_B$, but now $\widehat{\A}_R$ acts only on $\H_A$ while $\widehat{\B}_R$ acts only on $\H_B$. In particular, the full right boundary algebra $(\widehat{\A}_R \cup \widehat{\B}_R)''$ becomes simply $\A \otimes \B$ (with $\hat a_{R,i} \in \widehat{\A}_R$ acting as $\hat a_i \in \A$ and $\hat b_{R,i} \in \widehat{\B}_R$ acting as $\hat b_i \in \B$ respectively).

The leading perturbative correction to the correlation functions $C(t)$ is $O(\alpha)$ and is given by
\begin{align}
    \delta C = \ii \alpha &\braket{\Psi_A|\hat a_1 P_A \hat a_2  \dots |\Psi_A} \braket{\Psi_B|\hat b_1 P_B \hat b_2  \dots |\Psi_B} \nonumber\\
    &- \ii \alpha \braket{\Psi_A|\hat a_1  \hat a_2 P_A  \dots |\Psi_A} \braket{\Psi_B|\hat b_1 P_B \hat b_2  \dots |\Psi_B} \\&\qquad+ \dots \,. \nonumber
\end{align}
This is the usual single-graviton perturbative correction to out-of-time-ordered correlators close to the scrambling time, with a Lyapunov growth that saturates the chaos bound \cite{Maldacena:2015waa}. 

\subsection{Free product limit}
What about the limit $t \to + \infty$? Unlike the tensor product limit, to show that the modular-twisted product becomes a free product as $t \to +\infty$ will require some assumptions about the half-sided modular inclusions $\widetilde\A \subseteq \A$ and $\widetilde\B \subseteq \B$. Specifically, we will need to use the fact that the $*$-subalgebra $\widehat\A\subseteq\A$ defined in \eqref{eq:densesubalgebra} and the analogous $*$-subalgebra $\widehat\B\subseteq\B$ are s.o.t. dense. This will be important because, for this limit, we will choose the $*$-subalgebras $\widehat{\A}_R \subseteq \A_R$ and $\widehat{\B}_R \subseteq \B_R$ to be $\widehat\A_R = e^{\ii P_A P_B/2} \,\widehat\A\,e^{-\ii P_A P_B/2}$ and $\widehat\B_R = e^{-\ii P_A P_B/2} \,\widehat\B\,e^{\ii P_A P_B/2}$ respectively.

\subsubsection*{Modular cluster decomposition}
We first show that operators in $\widehat\A$ (and hence also $\widehat\B$) satisfy cluster decomposition when separated by a large modular translation.  To do so, it will be helpful to introduce notation where, given a partition $\pi$ of $\{1,2,\dots, n\}$ into subsets $I$, we write
\begin{align}\label{eq:partitioncorrelationnotation}
    \braket{\hat a_1 \hat a_2 \dots \hat a_n}_\pi = \prod_{I \in \pi} \braket{\Psi_A|\prod_{i \in I} \hat a_i |\Psi_A}\,.
\end{align}
Here (and elsewhere), the product over $i \in I$ is ordered so that $\hat a_i$ is to the left of $\hat a_j$ whenever $i<j$. Also, for $\hat a \in \widehat\A$, let $\hat a[s] = e^{\ii P_A s} \hat a e^{-\ii P_A s}$. (Note we are using square brackets here to specify a modular translation as opposed to a modular flow.) Cluster decomposition says, essentially, that 
\begin{align}\label{eq:heuristicclusterdecomp}
\braket{\hat a_1[s_1] \hat a_2[s_2] \dots \hat a_n[s_n]} \approx \braket{\hat a_1[s_1] \hat a_2[s_2] \dots \hat a_n[s_n]}_\pi
\end{align}
whenever each $s_i$ is only ``close'' to other $s_j$ associated to the same subset of $\pi$.

We can make \eqref{eq:heuristicclusterdecomp} somewhat more precise as follows. (Readers happy to accept \eqref{eq:heuristicclusterdecomp} as is without further clarification or derivation can safely skip to the next subsection.) Let the functions $s_i(\lambda)$ are chosen such that $\lim_{\lambda \to \infty} (s_i - s_j)$ exists in $\mathbb{R} \cup \{-\infty,\infty\}$ for all $i,j$. We want to show that
\begin{align}\label{eq:clusterdecompprecise}
    \lim_{\lambda \to \infty} \braket{\Psi_A| \hat a_1[s_1] \hat a_2[s_2] \dots \hat a_n[ s_n]|\Psi_A} = \lim_{\lambda \to \infty} \braket{ \hat a_1[s_1] \hat a_2[s_2] \dots \hat a_n[ s_n]}_\pi,
\end{align}
where the partition $\pi$ divides the operators into $I \subseteq \{1, \dots, n\}$ such that $\lim_{\lambda \to \infty} (s_i - s_j)$ is finite if and only if $i$ and $j$ are contained in the same subset $I$. In other words, the correlation function decomposes into pieces that only contain operators separated by a finite modular translation. 

In fact, we will show something slightly stronger, namely that 
\begin{align}\label{eq:wotlimit}
    \hat a_1[s_1] \hat a_2[s_2] \dots \hat a_n[ s_n] -  \prod_{I \in \pi} \braket{\Psi_A|\prod_{i \in I} a_i[s_i]|\Psi_A} \to 0
\end{align}
as $\lambda \to 0$ with respect to the weak operator topology (w.o.t.), so long as $s_i \to \infty$ for all $i$.  \eqref{eq:wotlimit} immediately implies \eqref{eq:clusterdecompprecise} since we can always simultaneously shift all the $s_i$ by any fixed function $f(\lambda)$ (and hence make $s_i \to \infty$ for all $i$) without changing the expectation value $\braket{a_1[s_1] \dots}$.

To prove \eqref{eq:wotlimit}, we first note that, for sufficiently large $\lambda$, $[\hat a_i[s_i], \hat a_j[s_j]] = 0$ whenever $i$ and $j$ are not in the same subset $I \in \pi$. Hence, we can always restrict to $\lambda$ large enough that
\begin{align}
    \hat a_1[s_1] \hat a_2[s_2] \dots \hat a_n[ s_n] =  \prod_{I \in \pi} \left[\prod_{i \in I} \hat a_i[s_i]\right]
\end{align}
where the product over $I \in \pi$ is ordered so that $I$ is to the left of $J$ if $s_i - s_j \to +\infty$ for $i \in I$ and $j \in J$. 

We then prove that this converges to the desired limit by induction on the number of subsets $I \in \pi$. The base case of $m = 0$ subsets is trivial. Now assume \eqref{eq:wotlimit} holds for $m$ subsets and consider the case with $(m + 1)$ subsets. Let $J \in \pi$ be the earliest subset and fix some $j_0 \in J$. We have
\begin{align}\nonumber
    \lim_{\lambda \to \infty} \braket{\Psi_A| \prod_{I \in \pi} \left[\prod_{i \in I} a_i[s_i]\right]|\Psi_A} &= \lim_{\lambda \to \infty} \braket{\Psi_A| \left(\prod_{I \neq J} \left[\prod_{i \in I} a_i[s_i - s_{j_0}]\right]\right) \prod_{j \in J} a_j[s_j - s_{j_0}] |\Psi_A} 
    \\&= \prod_{I \in \pi} \braket{\Psi_A|\prod_{i \in I} a_i[s_i]|\Psi_A}\,,\label{eq:expectationconverges}
\end{align}
where, in the first step, we used the fact that $P_A \ket{\Psi_A} = 0$ to shift all operator insertions by a modular translation of $s_{j_0}$ and, in the second step, we used the assumption of \eqref{eq:wotlimit} for the $m$ later subsets together with the fact that $\prod_{j \in J} a_j[s_j - s_{j_0}]$ converges in the strong operator topology as $\lambda \to \infty$.

Now let us return to considering the w.o.t. limit for the full set of $m+1$ subsets. Since the operator-norm unit ball is compact in the weak operator topology, any sequence of operators whose norm remains bounded as $\lambda \to \infty$ must contain a w.o.t. convergent subsequence. In our case, the norm of $\hat a_1[s_1]  \dots $ is bounded by $\prod_i\lVert \hat a_i\rVert$. Moreover, since $s_i \to \infty$ for all $i$, any limit point of $\hat a_1[s_1]  \dots $ must be contained in $\cap_t \Delta_A^{-\ii t} \widetilde\A \Delta_A^{\ii t}$. Since \eqref{eq:densesubalgebra} is s.o.t. dense in $\A$, this intersection contains only $c$-numbers. Finally, \eqref{eq:expectationconverges} tells us that the only $c$-number that can be a w.o.t. limit point of $\hat a_1[s_1]  \dots $ is the right-hand side of \eqref{eq:wotlimit}. The desired result \eqref{eq:wotlimit} therefore follows immediately.

\subsubsection*{Free product correlators}
We are now ready to understand the limit of correlation functions $C(t)$ as $t \to +\infty$. First, note that
\begin{align}
    C(t) = \frac{1}{\alpha^n} \int dy_1 \dots dy_n \left\langle\hat a_1 \Pi_A\left(\frac{y_1}{\alpha}\right)\hat a_2\Pi_A\left(\frac{y_2}{\alpha}\right)\dots \right\rangle \braket{\hat b_1[y_1] \dots\hat b_{n}[y_n] }\,,
\end{align}
where $dy\,\Pi_A(y)$ is the projection-valued measure for $P_A$ as before. Since
\begin{align}
    \Pi_A(y)  = \delta(y - P_A) = \frac{1}{2 \pi}  \int dx e^{ix(P_A-y)}\,, \label{eq:deltareplacement}
\end{align}
we have
\begin{align}\label{eq:douglaszhenbinformula}
     C(t) =  \int \frac{dx_1 \dots dx_n dy_1 \dots dy_n}{(2 \pi \alpha)^n} \braket{\hat a_1[x_1] \dots} \braket{\hat b_1[y_1]  \dots} e^{-\frac{\ii}{\alpha}\sum_iy_i(x_{i+1} - x_i)}\,.
\end{align}
The sum in the exponent is from $1$ to $n$, with $x_{n+1} = 0$. When using \eqref{eq:deltareplacement} to substitute for $\Pi_A(y_k)$, we have replaced $x$ by $x_{k+1} - x_{k}$. This is a generalization to $2n$-point OTOCs of Eq. 2.7 in \cite{Stanford:2021bhl}.

It will be helpful to decompose the correlation function  $\braket{\hat a_1 \hat a_2 \dots \hat a_n}$ (and likewise $\braket{\hat b_1 \hat b_2 \dots \hat b_n}$) into connected correlation functions $\braket{\hat a_1 \dots }_{\rm conn}$ via the usual formula
\begin{align}\label{eq:connecteddef}
    \braket{\hat a_1 \hat a_2 \dots \hat a_n} = \sum_{\pi} \braket{\hat a_1 \hat a_2 \dots \hat a_n}_{\pi,{\rm conn}}\,,
\end{align}
where the sum is over all partitions of $\{1,2,\dots n\}$ with $\braket{\hat a_1  \dots}_{\pi,{\rm conn}}$ defined analogously to \eqref{eq:partitioncorrelationnotation}. (The formula \eqref{eq:connecteddef} uniquely defines $\braket{\hat a_1 \hat a_2 \dots \hat a_n}_{\rm conn}$ by taking a M\"{o}bius inversion.)

\begin{figure}[t]
\centering
\begin{subfigure}{0.45\linewidth}

    \includegraphics[width = 7.7cm]{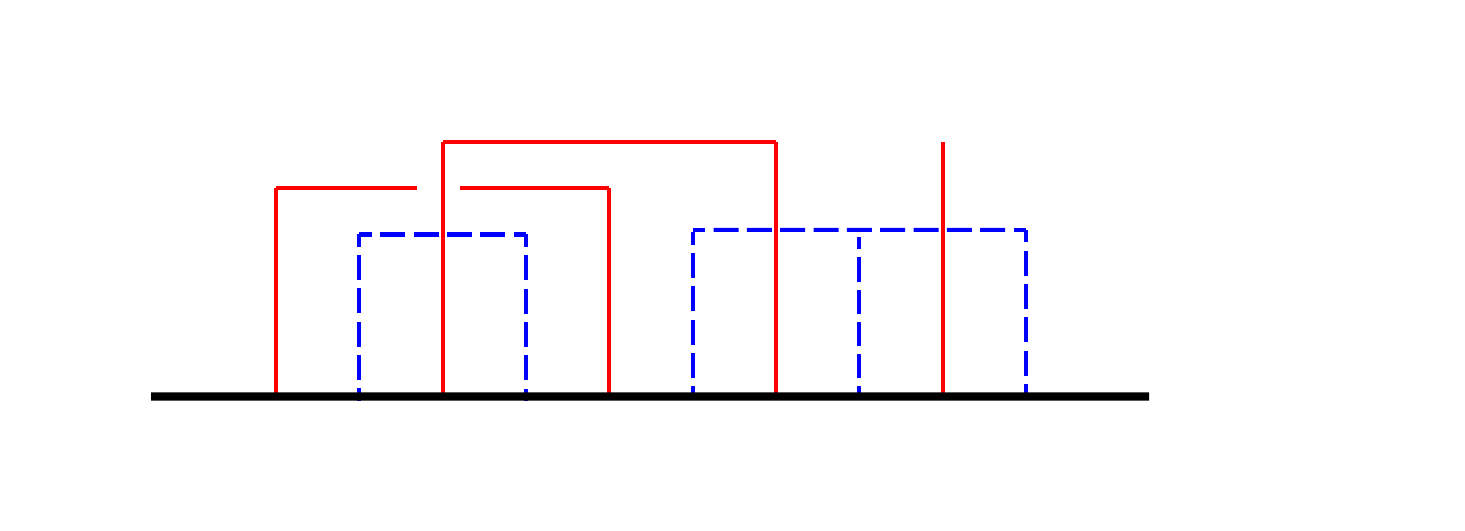}
    \centering
    \caption{}\label{fig:interlaced}
\end{subfigure}
\begin{subfigure}{0.45\linewidth}
    \centering
    \includegraphics[width=7.7cm]{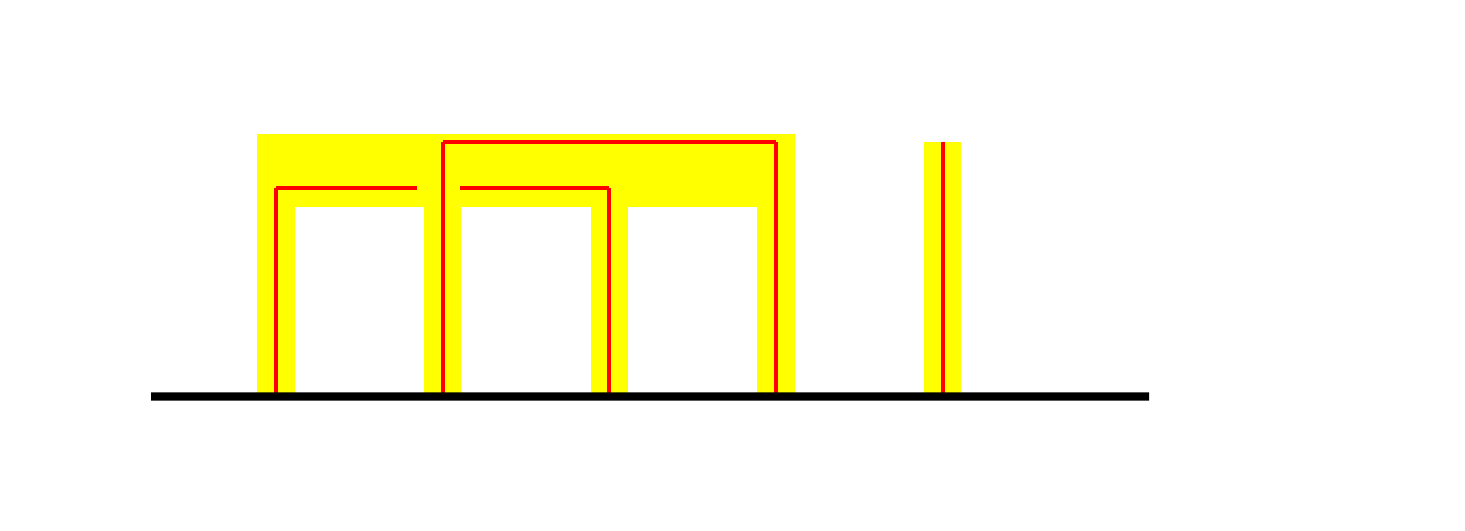}
    \caption{}\label{fig:noncrossingclosure}
\end{subfigure}
\begin{subfigure}{0.45\linewidth}
    \centering
    \includegraphics[width=7.7cm]{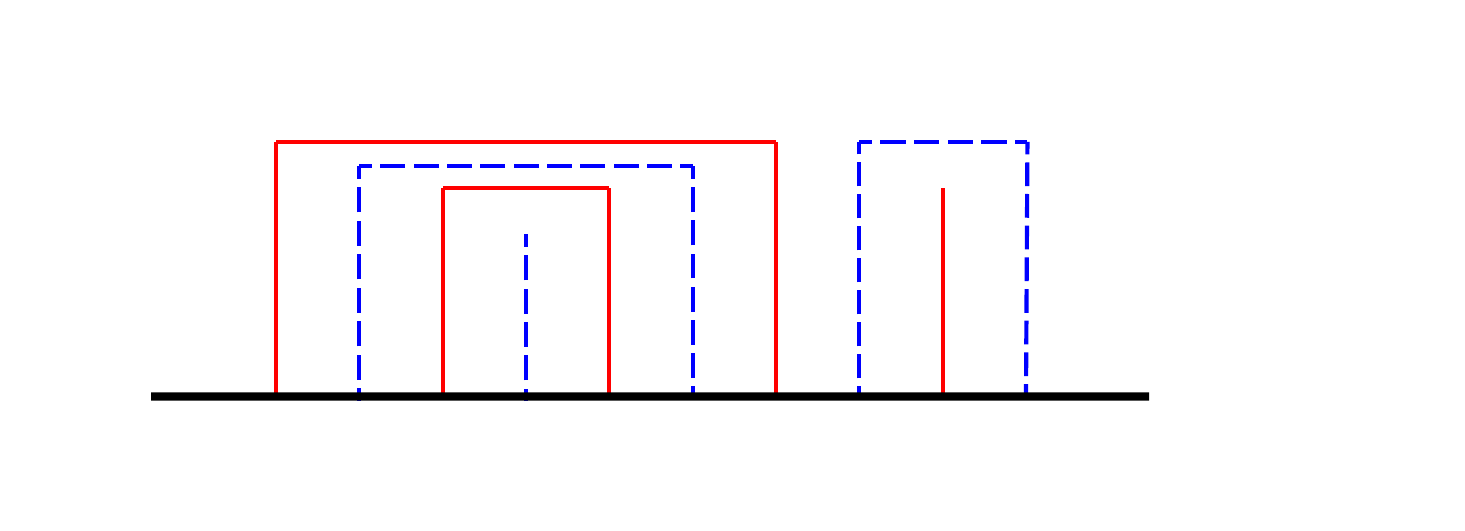}
    \caption{}\label{fig:kreweras}
\end{subfigure}
\caption{(a) Two partitions $\pi_A = \{\{1,3\},\{2,4\},\{5\}$ (solid red) and $\pi_B = \{\{1,2\},\{3,4,5\}\}$ (dashed blue) can be represented graphically by interlacing them. (b) Given a partition $\pi = \{\{1,3\},\{2,4\},\{5\}\}$ (red), the noncrossing closure $\bar\pi = \{\{1,2,3,4\},\{5\}\}$ (yellow) is found by grouping all partitions that cross one another. (c) Given a noncrossing partition $\pi = \{\{1,4\},\{2,3\},\{5\}\}$, the Kreweras complement $K(\pi)= \{\{1,3\},\{2\},\{4,5\}\}$ is the coarsest partition that can be interlaced with $\pi$ without crossing it.}\label{fig:graphicpartitions}
\end{figure}

Let us expand \eqref{eq:douglaszhenbinformula} using \eqref{eq:connecteddef} as
\begin{align}
    C(t) = \sum_{\pi_A, \pi_B} C(t)_{\pi_A, \pi_B}
\end{align}
 with 
 \begin{align}
     C(t)_{\pi_A, \pi_B} = \int \frac{dx_1 \dots dx_n dy_1 \dots dy_n}{(2 \pi \alpha)^n} \braket{\hat a_1[x_1] \dots}_{\pi_A, {\rm conn}} \braket{\hat b_1[y_1]  \dots}_{\pi_B, {\rm conn}} e^{-\frac{\ii}{\alpha}\sum_iy_i(x_{i+1} - x_i)}\,.
 \end{align}
It will be helpful to represent the partitions $\pi_A$ and $\pi_B$ by interlacing them on a line, as shown in Figure \ref{fig:interlaced}. 
We will show that the contribution to \eqref{eq:douglaszhenbinformula} vanishes for any pair of partitions $\pi_A$ and $\pi_B$ that cannot be drawn in this way without a subset of $\pi_A$ crossing a subset of $\pi_B$. (Crossings between two subsets of $\pi_A$ or two subsets of $\pi_B$ are allowed.) Moreover, the contributions from allowed pairs of partitions will turn out to be
\begin{align}
    \lim_{t \to +\infty} C(t)_{\pi_A,\pi_B} = \braket{\hat a_1 \dots}_{\pi_A, {\rm conn}} \braket{\hat b_1 \dots}_{\pi_B, {\rm conn}} \,,
\end{align}
so that
\begin{align}\label{eq:summutuallynoncrossing}
    \lim_{t \to +\infty}& C(t) = \sum_{\substack{\pi_A, \pi_B \\{\rm mutually} \\{\rm noncrossing}}} \braket{\hat a_1 \dots \hat a_n}_{\pi_A,{\rm conn}} \braket{\hat b_1 \dots \hat b_n}_{\pi_B, {\rm conn}} \,.
\end{align}

The formula \eqref{eq:summutuallynoncrossing} can be rewritten in a slightly cleaner form as follows. For any partition $\pi$, we define the noncrossing closure $\bar\pi$ to be the finest noncrossing partition such that $\pi$ is a refinement of $\bar \pi$.\footnote{ A partition $\bar \pi$ is noncrossing if it can be depicted graphically as in Figure \ref{fig:graphicpartitions} without any intersections between different subsets of $\bar\pi$. A partition $\pi_1$ is a refinement of $\pi_2$ if each block of $\pi_1$ is contained in a block of $\pi_2$.  For example, $\{\{1\},\{2\}\}$ is a refinement of $\{1,2\}$. See Definition 9.14 of \cite{Nica_Speicher_2006}.} Graphically, $\bar \pi$ can be found by grouping together all subsets of $\pi$ that intersect, as shown in Figure \ref{fig:noncrossingclosure}. Given a noncrossing partition $\pi$, we can define the free cumulant
\begin{align}
    \braket{\hat a_1 \hat a_2 \dots \hat a_n}_{\pi, {\rm free}} = \sum_{\pi': \,\bar{\pi}' = \pi} \braket{\hat a_1 \hat a_2 \dots \hat a_n}_{\pi',{\rm conn}}\,,
\end{align}
where the sum is over all partitions $\pi'$ whose noncrossing closure is $\pi$. This satisfies an analogous formula to \eqref{eq:connecteddef} except that the sum over all partitions is replaced by a sum over only noncrossing partitions. Finally, given a noncrossing partition $\pi$, the Kreweras complement $K(\pi)$ is the coarsest partition that is mutually noncrossing when interlaced with $\pi$. It can be found graphically by connecting all interlaced points that are not separated by the partition $\pi$; see Figure \ref{fig:kreweras}. 

Two partitions $\pi_A$ and $\pi_B$ are mutually noncrossing if and only if $\pi_B$ is a refinement of the Kreweras complement $K(\bar\pi_A)$. As a result, we have
\begin{align}\label{eq:sumnoncrossing}
    \lim_{t \to +\infty} C(t) &= \sum_{\substack{\pi_A \\{\rm noncrossing}}} \braket{\hat a_1 \dots \hat a_n}_{\pi_A,{\rm free}} \braket{\hat b_1 \dots \hat b_n}_{K(\pi_A)} 
    \\&= \sum_{\substack{\pi_B \\{\rm noncrossing}}} \braket{\hat a_1 \dots \hat a_n}_{K(\pi_B)} \braket{\hat b_1 \dots \hat b_n}_{\pi_B,{\rm free}}\,,\nonumber
\end{align}
which is a standard formula for correlation functions of freely independent variables (see e.g. Theorem 14.4 of \cite{Nica_Speicher_2006}) and shows that indeed the $t \to - \infty$ algebra is a free product. 

To see this last statement, note that whenever $\pi_A$ and $\pi_B$ are mutually noncrossing, they must contain at least two singleton sets between them. This is because they are refinements of (respectively) $\bar \pi_A$ and $K(\bar \pi_A)$ and it is easy to prove that the number of subsets in a noncrossing partition $\pi$ plus the number of subsets in $K(\pi)$ is exactly $(n+1)$. Moreover, if there are only two singleton subsets, they cannot be $\{1\} \in \pi_A$ and $\{n\}\in \pi_B$ because then $\pi_A$ restricted to $\{2, \dots, n\}$ and $\pi_B$ restricted to $\{1, \dots,(n-1)\}$ would be mutually noncrossing partitions containing no singleton sets. It follows that \eqref{eq:summutuallynoncrossing} vanishes whenever $\braket{\hat a_i} = \braket{\hat b_j} = 0$ (except possibly for $\hat a_1$, $\hat b_n$ or both which may be equal to the identity). This is the defining property of a free product von Neumann algebra.

\subsubsection*{Warm-up: \eqref{eq:summutuallynoncrossing} for four-point functions}
It remains then only to derive the formula \eqref{eq:summutuallynoncrossing}. As a warm-up, we consider the four-point correlator, i.e. $n=2$, and derive \eqref{eq:summutuallynoncrossing} from \eqref{eq:douglaszhenbinformula}. This was already done in \cite{Stanford:2021bhl}, although we will expand on the details of the derivation somewhat. There are four possible pairs of partitions to consider. For $\pi_A=\pi_B=\{\{1\},\{2\}\}$, the vacuum invariance under modular translations means that the matter correlators factor out of the integrals and we have
\begin{align}
    \lim_{t \to +\infty} C(t)_{\pi_A,\pi_B} &= \braket{\hat a_1}\braket{\hat a_2} \braket{\hat b_1}\braket{\hat b_2} \lim_{\alpha \to \infty}\int \frac{dx_1dx_2 dy_1dy_2}{(2 \pi \alpha)^2} e^{-\frac{\ii}{\alpha} [y_1(x_2 - x_1) - y_2x_2]} \nonumber\\
    &= \braket{\hat a_1}\braket{\hat a_2} \braket{\hat b_1}\braket{\hat b_2}\,.
\end{align}
For $\pi_A=\{\{1\},\{2\}\}$ and $\pi_B=\{1,2\}$, we write $y_1=Y+y'$ and $y_2=Y-y'$ where
\begin{align}
    Y =\frac12(y_1+y_2)\,,\quad y' =\frac12(y_1-y_2) \,.
\end{align}
The matter correlators are independent of $x_1$, $x_2$ and $Y$, so these integrals can be done explicitly and we obtain
\begin{align}
    \lim_{t \to +\infty} C(t)_{\pi_A,\pi_B} &= \braket{\hat a_1}\braket{\hat a_2} \lim_{\alpha \to \infty}\int \frac{2dx_1dx_2 dYdy'}{(2 \pi \alpha)^2} e^{-\frac{\ii}{\alpha} [-x_1Y + (2x_2-x_1)y']} \braket{\hat b_1[y'] \hat b_2}_{\rm conn} \nonumber\\
    &= \braket{\hat a_1}\braket{\hat a_2} \int dy' \delta(y') \braket{\hat b_1[y'] \hat b_2}_{\rm conn}= \braket{\hat a_1}\braket{\hat a_2} \braket{\hat b_1\hat b_2}_{\rm conn}\,.
\end{align}
Similarly, for $\pi_A=\{1,2\}$ and $\pi_B=\{\{1\},\{2\}\}$, we write $x_1=X+x'$ and $x_2=X-x'$ where
\begin{align}
    X =\frac12(x_1+x_2)\,,\quad x' =\frac12(x_1-x_2) \,.
\end{align}
We then have
\begin{align}
    \lim_{t \to +\infty} C(t)_{\pi_A,\pi_B} &= \braket{\hat b_1}\braket{\hat b_2} \lim_{\alpha \to \infty}\int \frac{2dX dx' dy_1dy_2}{(2 \pi \alpha)^2} e^{-\frac{\ii}{\alpha} [-Xy_2 + x'(y_2-2y_1)]} \braket{\hat a_1[x'] \hat a_2}_{\rm conn} \nonumber\\
    &= \braket{\hat a_1\hat a_2}_{\rm conn} \braket{\hat b_1}\braket{\hat b_2}\,.
\end{align}
Finally, for $\pi_A=\pi_B=\{1,2\}$, we find
\begin{align}
    \lim_{t \to +\infty} C(t)_{\pi_A,\pi_B} &= \lim_{\alpha \to \infty}\int \frac{4dX dx' dYdy'}{(2 \pi \alpha)^2} e^{-\frac{\ii}{\alpha} [-XY + Xy' - x'Y - 3x'y']} \braket{\hat a_1[x'] \hat a_2}_{\rm conn} \braket{\hat b_1[y'] \hat b_2}_{\rm conn} \nonumber\\
    &= \lim_{\alpha \to \infty}\int \frac{4dx'dy'}{2 \pi \alpha} e^{\frac{\ii}{\alpha} [4x'y']} \braket{\hat a_1[x'] \hat a_2}_{\rm conn} \braket{\hat b_1[y'] \hat b_2}_{\rm conn} \nonumber\\
    &= \left[\lim_{\alpha\to \infty} \frac{1}{\alpha}\right] \int \frac{4dx'dy'}{2 \pi } \braket{\hat a_1[x'] \hat a_2}_{\rm conn} \braket{\hat b_1[y'] \hat b_2}_{\rm conn} = 0 \,.
\end{align}
In the third step, we used the fact that $\braket{\hat a_1[x'] \hat a_2}_{\rm conn}$ and $\braket{\hat b_1[y'] \hat b_2}_{\rm conn}$ vanish by cluster decomposition at large $x'$, $y'$ in order to set the exponent equal to zero in the limit $\alpha \to \infty$. After doing so, the integrals over $x'$ and $y'$ give projection-valued measures $\Pi_A(0)$ and $\Pi_B(0)$ respectively; these give a finite answer when inserted in connected correlation functions since the divergent vacuum contribution is cancelled out by subtracting the product of one-point functions. 

Putting everything together, we recover \eqref{eq:summutuallynoncrossing} for $n=2$, i.e. that the correlator is a sum over the possible mutually noncrossing partitions:
\begin{align}
    \lim_{t \to +\infty} C(t) = \braket{\hat a_1}\braket{\hat a_2} \braket{\hat b_1 \hat b_2}_{\rm conn} + \braket{\hat a_1 \hat a_2}_{\rm conn} \braket{\hat b_1}\braket{\hat b_2} + \braket{\hat a_1}\braket{\hat a_2} \braket{\hat b_1}\braket{\hat b_2} \,.
\end{align}

\subsubsection*{Derivation of \eqref{eq:summutuallynoncrossing} for arbitrary $n$}
The derivation for general $n$ follows the same spirit as the $n=2$ case considered above, except that the details are somewhat messier. The key point is that, as a consequence of \eqref{eq:heuristicclusterdecomp}, connected correlation functions vanish whenever the separation between any two operator insertions becomes parametrically large. As a result, if for each subset $I_a \in \pi_A$ (or subset $I_b \in \pi_B$), we write 
\begin{align}\label{eq:Xadef}
    x_i &= X_a + x_i' \qquad \qquad X_a = \mathbb{E}_{j \in I_a} x_j \\
    y_i &= Y_b + y_i' \qquad \qquad Y_b = \mathbb{E}_{j \in I_b} y_j \,, \label{eq:Ybdef}
\end{align}
then the matter correlation functions will vanish unless $x_i' = O(1)$ for all $i$. (In contrast, the variables $X_a$ may be arbitrarily large.) In terms of these new variables, we have
\begin{align}
    \prod_{i \in I_a} dx_i = dX_a \prod_{i \in I_a} dx_i'\, \delta\left(\mathbb{E}_{j \in I_a} x_j'\right) \qquad \text{and} \qquad\prod_{i \in I_b} dy_i = dY_b \prod_{i \in I_b} dy_i'\, \delta\left(\mathbb{E}_{j \in I_b} y_j'\right) \,.
\end{align}
We also have
\begin{align} \label{eq:exponentexpand}
    \sum_iy_i(x_{i+1} &- x_i) = \vec{y}^T M \vec{x} = \vec{Y}^T W^T_{\pi_B} M V_{\pi_A} \vec{X} + \vec{y'}{}^T M V_{\pi_A} \vec{X} + \vec{Y}^T W^T_{\pi_B} M \vec{x'} + \vec{y'}{}^T M \vec{x'} \nonumber\\
    &= Y_b W^{ib}_{\pi_B} M_{ij} V^{ja}_{\pi_A} X_a + y'{}^i M_{ij} V^{ja}_{\pi_A} X_a + Y_b W^{ib}_{\pi_B} M_{ij} x'{}^j + y'{}^i M_{ij} x'{}^j\,,
\end{align}
where
\begin{align}
    M_{ij} = \delta_{i+1,j} - \delta_{i,j} \,,
\end{align}
while
\begin{align}
 V_{\pi_A}^{j a} = \begin{cases}
    1 & \text{if } j \in I_a \in \pi_A \\
    0   & \text{otherwise }
\end{cases} \qquad\text{and}\qquad W_{\pi_B}^{i b} = \begin{cases}
    1 & \text{if } i \in I_b \in \pi_B \\
    0   & \text{otherwise }
\end{cases}\,.
\end{align}
In the $\alpha \to \infty$ limit, the last term in \eqref{eq:exponentexpand} is $O(1)$ and so can be ignored when writing the exponent in \eqref{eq:douglaszhenbinformula}. 

Putting everything together, \eqref{eq:douglaszhenbinformula} becomes
\begin{align}\label{eq:douglaszhenbinformula2}\nonumber
    \lim_{t \to +\infty} C_{\pi_A,\pi_B}(t) =  \lim_{\alpha \to \infty} \int \prod_{a, b, i, j}&\frac{dX_a dY_b dx'_i dy'_j}{(2 \pi \alpha)^n}  e^{\frac{\ii}{\alpha}\left(\vec{Y}^T W_{\pi_B}^T M V_{\pi_A} \vec{X} + \vec{y'}{}^T M V_{\pi_A} \vec{X} + \vec{Y}^T W_{\pi_B}^T M \vec{x'}\right)} \\&\qquad\times\braket{\hat a_1[x_1'] \dots}_{\pi_A,{\rm conn}} \braket{\hat b_1[y_1']  \dots}_{\pi_B,{\rm conn}}\,.
\end{align}
Because the modular translation leaves the vacuum invariant, the integrals over $X_a$ and $Y_b$ no longer depend on the matter correlation functions and can be carried out explicitly. Each integral over a variable $X_a$ gives a delta function involving a linear combination of $Y_b$ and $y'_j$, together with a factor of $\alpha$. In total, those integrals leads to an overall factor of $\alpha^{n_A}$ where $n_A$ is the number of subsets in $\pi_A$. 

After integrating over the $X_A$, we may be left with some remaining integrals over $Y_b$ that were not fixed (in terms of $y'_j$) by the delta functions. More precisely, these remaining integrals are over the cokernel of the map $W^T_{\pi_B} M V_{\pi_A}$. Each such integral gives delta function involving a linear combination of $x_i'$ and a factor of $\alpha$. This second step therefore gives an overall factor of $\alpha^{n_B - r_{AB}}$ where $n_B$ is the number of subsets in $\pi_B$ and $r_{AB}$ is the rank of the matrix $W^T_{\pi_B} M V_{\pi_A}$. Finally, we are left with an $\alpha$-independent integral over any linear combinations of $x'_i$ or $y'_j$ that have not yet been fixed by delta functions.

Given the factor of $\alpha^n$ in the denominator of \eqref{eq:douglaszhenbinformula2}, the analysis above tells us that $C_{\pi_A,\pi_B}(t)$ vanishes as $t \to +\infty$ for any $\pi_A, \pi_B$ where\begin{align}\label{eq:disallowedpartitions}
    n_A + n_B - r_{AB} < n\,.
\end{align}
In the case $n=2$, this occurs only when $\pi_A = \pi_B = \{1,2\}$, as shown in Table \ref{tab:}.
\begin{table}[t]
    \centering
    \begin{tabular}{cccccccc}
        $\pi_A$ & $\pi_B$ & $V$ & $W$ & $W^TMV$ & $r_{AB}$ & $n_A + n_B-r_{AB}$ \\ \hline\hline
        \{12\} & \{12\} & $\begin{pmatrix} 1\\1 \end{pmatrix}$ & $\begin{pmatrix} 1\\1 \end{pmatrix}$ & $-1$ & $1$ & $1+1-1=1$ \\
        \{12\} & \{1\}\{2\} & $\begin{pmatrix} 1\\1 \end{pmatrix}$ & $I_2$ & $\begin{pmatrix} 0\\-1 \end{pmatrix}$ & $1$ & $1+2-1=2$ \\
        \{1\}\{2\} & \{12\} & $I_2$ & $\begin{pmatrix} 1\\1 \end{pmatrix}$ & $\begin{pmatrix} -1&0 \end{pmatrix}$ & $1$ & $2+1-1=2$ \\
        \{1\}\{2\} & \{1\}\{2\} & $I_2$ & $I_2$ & $\begin{pmatrix} -1&1\\0&-1 \end{pmatrix}$ & $2$ & $2+2-2=2$ \\
    \end{tabular}
    \caption{The matrices $V$, $W$ and $W^T MV$ (as well as the rank $r_{AB}$ of the latter) in the case $n=2$. We see that $n_A + n_B - r_{AB} = n$ except when $\pi_A = \pi_B = \{1,2\}$.}
    \label{tab:}
\end{table}

For general $n$, we first show that \eqref{eq:disallowedpartitions} implies $C_{\pi_A, \pi_B}(t) \to 0$ unless the partitions $\pi_A, \pi_B$ are mutually noncrossing. Let the partition $\pi_A'$ be a refinement of $\pi_A$. We can write $V_{\pi_A} = V_{\pi_A'} V_{\pi_A \to \pi_A'}$ where $V_{\pi_A \to \pi_A'}^{a' a} = 1$ if $a' \subseteq a$ and $V_{\pi_A \to \pi_A'}^{a' a} = 0$ otherwise. Composition with $V_{\pi_A \to \pi_A'}$ decreases the rank of $W^T_{\pi_B} M V_{\pi_A'}$ by at most $n_A' - n_A$ (where $n_A'$ is the number of subsets in $\pi_A'$) -- or by less if the kernel of $W^T_{\pi_B} M V_{\pi_A}$ is not contained in the image of $\tilde V_{\pi_A' \to \pi_A}$. It follows that, to see whether \eqref{eq:disallowedpartitions} is satisfied for some pair of partitions $\pi_A, \pi_B$, it suffices to check the case $\pi_A',\pi_B$ where $\pi_A'$ is any refinement of $\pi_A'$. An exactly analogous argument  (with $V_{\pi_A}$ replaced by $W_{\pi_B}$) shows that we can also replace $\pi_B$ by any refinement $\pi_B'$.

Unless $\pi_A, \pi_B$ are mutually noncrossing, we can find refinements $\pi_A'$, $\pi_B'$, where a) $\pi_A'$ and $\pi_B'$ each consist of singletons except for a single pair $\{i_A, j_A\}$ and $\{i_B, j_B\}$ respectively and b) the two pairs cross when interlaced. More explicitly, for $i_A < j_A$, we demand that $i_B \in [i_A, j_A-1]$ and $j_B \not\in [i_A, j_A-1]$. 

Using \eqref{eq:disallowedpartitions}, we conclude that $C_{\pi_A, \pi_B} \to 0$ as $\alpha \to \infty$ unless $W^T_{\pi_B'} M V_{\pi_A'}$ has rank at most $(n-2)$. But the $(n-1)$-dimensional image of $M V_{\pi_A}$ consists of vectors $x_i$ satisfying
\begin{align}
    \sum_{i = i_A}^{j_A -1} x_i = 0 \,, 
\end{align}
while the kernel of $W_{\pi_B}^T$ consists of vectors with $x_{i_B} = -x_{j_B}$ and $x_i = 0$ otherwise. These subspaces do not intersect if $\{i_A, j_A\}$ and $\{i_B, j_B\}$ cross and so the image of $W^T_{\pi_B'} M V_{\pi_A'}$ is $(n-1)$-dimensional. We conclude that the contribution to \eqref{eq:douglaszhenbinformula2} comes only from mutually noncrossing partitions $\pi_A, \pi_B$.

Now let us assume that $\pi_A$ and $\pi_B$ are indeed mutually noncrossing. For this part of the argument, it will prove convenient to replace $\pi_A$ and $\pi_B$ in the definitions of $X_a$, $x_i'$, $Y_b$ and $y_i'$ given in \eqref{eq:Xadef} and \eqref{eq:Ybdef} by $\bar{\pi}_A$ and $K(\bar \pi_A)$ respectively. If we do this, we can no longer assume that matter correlators vanish unless the parameters $x_i'$ or $y_j'$ are $O(1)$, but we can still use the fact that those correlators are independent of $X_a$ and $Y_b$ to carry out the integrals over the $X_a$ and $Y_b$ variables explicitly. Also, it is easy to check that, with these new definitions, we have
\begin{align}
    Y_b W^{ib}_{K(\bar \pi_A)} M_{ij} V_{\pi_A}^{ja} X_a = X_{a_1}Y_{b_n} \,,
\end{align}
where $a_1 \in \bar \pi_A$ is the subset of $\pi_A$ that contains the element $1$, while $b_n \in K(\bar \pi_A)$ is the subset that contains the element $n$. If $\bar n_{A}$ is the number of subsets in $\bar \pi_A$, this leaves $(\bar{n}_A-1)$ remaining $X_a$-integrals that each give a delta function involving only linear combinations of $y_j'$. If there are $\bar{n}_B$ subsets in $K(\bar \pi_A)$, then there are $n - \bar{n}_B$ independent $y_j'$-variables. But, since $\bar n_A + \bar n_B = n + 1$, that means that all the $y_j'$-variables (along with $Y_{b_n}$) are fixed to be zero by the integral over $X_A$-variables.\footnote{We know that the delta functions (both here and in the $Y_b$-integrals) must all give independent constraints, since the integral in \eqref{eq:nomatter} is finite.} We are left with
\begin{align}\label{eq:douglaszhenbinformula3}
    \lim_{t \to +\infty} C_{\pi_A, \pi_B}(t) =  A_{\bar \pi_A} \lim_{\alpha \to \infty}\int \prod_{b, i}&\frac{dY_b dx'_i }{(2 \pi \alpha)^{\bar{n}_B - 1}}  e^{\frac{\ii}{\alpha}\vec{Y}^T W_{\pi_B}^T M \vec{x'}}\\&\times \braket{\hat a_1[x_1']\hat a_2[x_2'] \dots}_{\pi_A,{\rm conn}} \braket{\hat b_1\hat b_2  \dots}_{\pi_B,{\rm conn}}\,,\nonumber
\end{align}
where $A_{\bar \pi_A}$ is a $O(1)$ constant that could have been introduced by the $X_A$-integrals. 

The remaining $(\bar n_B -1)$ $Y_b$-integrals give delta functions that fix all the  $x_i'$-variables to zero (since again $n - \bar n_A = \bar n_B -1$). They may also give an $O(1)$ constant $B_{\bar \pi_A}$, so we find that  
\begin{align}\label{eq:douglaszhenbinformula4}
    \lim_{t \to +\infty} C_{\pi_A,\pi_B}(t) =  A_{\bar \pi_A}B_{\bar \pi_A} \braket{\hat a_1\hat a_2 \dots}_{\pi_A,{\rm conn}} \braket{\hat b_1\hat b_2  \dots}_{\pi_B,{\rm conn}}\,,
\end{align}
for any mutually noncrossing pair of partitions $\pi_A$, $\pi_B$. However, it is easy to check that $A_{\bar \pi_A} B_{\bar \pi_A} = 1$ since by identical arguments
\begin{align}\label{eq:nomatter}
    1 = \int \frac{dx_1 \dots dx_n dy_1 \dots dy_n}{(2 \pi \alpha)^n} e^{\frac{\ii}{\alpha}\vec{y}^TM\vec{x}} = A_{\bar\pi_A} B_{\bar\pi_A}\,,
\end{align}
for any partition $\bar \pi_A$.

\section{Entanglement entropy and the crossed product}\label{sec:typeii}

\subsection{Crossed product by the modular automorphism group}

In this section, we extend the Hilbert space by including the Hamiltonian in the large $N$ algebra. We first briefly review the construction in \cite{Chandrasekaran:2022eqq} of the crossed product algebra that describes the large $N$ limit of the microcanonical ensemble.

The thermofield double state
\begin{equation}
    |{\rm TFD}\> = \sum_i e^{-\beta E_i/2} |E_i\>_L |E_i\>_R 
\end{equation}
 is a purification of the canonical ensemble $\rho = e^{-\beta H_R}$. In the large $N$ limit, the thermal expectation value $\<H_R\>_\beta \sim\O(N^2)$ of the boundary Hamiltonian diverges. Moreover, even if we subtract that divergence, the fluctuations
\begin{equation}
    \< (H_R-\<H_R\>_\beta)^2\>_\beta \sim \O(N^2)
\end{equation}
will still diverge. Instead, an operator with finite fluctuations is the rescaled Hamiltonian
$(H_R-\<H_R\>)/N$.
In the limit $N \to \infty$,
\begin{align}
   \frac{1}{N}[(H_R-\<H_R\>),a] = -\frac1N\ii\p_t a\to 0 \qquad \forall a \in \A \,.
\end{align}
So, this rescaled Hamiltonian is central in the large $N$ limit of the canonical ensemble, as described in Section \ref{sec:review}. If we included it in our large $N$ algebra, we would obtain a tensor product of the Type III$_1$ factor $\A$ with an infinite-dimensional center consisting of functions of the rescaled Hamiltonian.

Things work quite differently, however, in the microcanonical ensemble, where energy fluctuations are kept finite as $N \to \infty$. Suppose we start from a microcanonical version of the thermofield double
\begin{equation}\label{eq:microTFD}
    |\widetilde{\rm TFD}\> = e^{-S(E_0)/2} \sum_i e^{-\beta (E_i-E_0)/2} f(E_i-E_0) |E_i\>_L |E_i\>_R \,,
\end{equation}
where $f(E-E_0)$ is an arbitrary smooth, invertible and normalizable function. (The large $N$ algebra and Hilbert space we construct will end up independent of the initial choice of $f$.) At large $N$, the state $ |\widetilde{\rm TFD}\>$ has the same correlation functions as $|{\rm TFD}\>$ for single-trace operators other than the Hamiltonian and hence leads to the same large $N$ algebra $\A$. However, it has finite fluctuations of the renormalized Hamiltonian 
\begin{align}
    h_R=H_R-E_0
\end{align} 
with no rescaling required. 

Unlike in the canonical ensemble, the operator $h_R$ is not central; instead, it satisfies
\begin{equation}\label{eq:hRflow}
    [h_R,a] = -\ii\p_t a = \O(1)\,.
\end{equation}
Including $h_R$ in the large $N$ algebra leads to an algebra called the modular crossed product $\A\rtimes\R$. It follows from \eqref{eq:hRflow} that
\begin{equation}\label{eq:hamdifference}
    \beta h_R = \beta h_L - \log \Delta_A\,,
\end{equation}
where $h_L$ is an operator that commutes with $\A$. Since, at finite $N$, we have
\begin{align}
    - \log \Delta_A = \beta (H_R -H_L)\,,
\end{align}
with $H_L$ the left boundary Hamiltonian, we can identify $h_L$ with the large $N$ limit of $H_L - E_0$. This operator has purely continuous spectrum in the $N \to \infty$ limit, so including it leads to the large $N$ Hilbert space
\begin{align}
\H_A^{\rm cr} = \H_A \otimes L^2(\R)
\end{align}
with $h_L$ acting as the position operator on $L^2(\R)$. The algebra generated by $h_R$ and $\A$ (the latter acting purely on $\H_A$) is, by definition, the crossed product $\A\rtimes\R$. 

From a bulk perspective, $h_L$ (resp. $h_R$) is the renormalized left (resp. right) ADM Hamiltonian, while $\log \Delta_\Psi$ is the boost Hamiltonian of (both left- and right-moving) bulk matter fields. The momentum conjugate to $h_L$ on $L^2(\R)$ describes the timeshift between the left and right boundary: the uncertainty principle means that states with finite fluctuations in $h_L$ (i.e. states in the microcanonical ensemble Hilbert space) also have finite fluctuations in this timeshift. See \cite{Chandrasekaran:2022eqq} for further details.

To add the renormalized Hamiltonian $h_R$ to the scrambling algebra constructed in Section \ref{sec:main}, we go through an essentially identical procedure to the one above. Recall that
\begin{equation}
    \log\Delta_\Psi = \log \Delta_A+\log \Delta_B \,.
\end{equation}
Since $[\log\Delta_\Psi, P_A P_B] = 0$, we therefore have
\begin{align}\label{eq:hpsiflow}
    [\log\Delta_\Psi, a_R] = \ii\beta \partial_t a_R \,, \qquad [\log\Delta_\Psi, b_R] = \ii\beta \partial_t b_R
\end{align}
for any $a_R \in \A_R$ and $b_R \in \B_R$.

If, as above, we construct our large $N$ algebra starting from the microcanonical TFD state \eqref{eq:microTFD}, the renormalized Hamiltonian $h_R = H_R - E_0$ again has a finite large $N$ limit that satisfies
\begin{align}\label{eq:hRscramblingflow}
    [h_R, a_R] = -\ii \partial_t a_R \,, \qquad [h_R, b_R] = -\ii \partial_t b_R\,.
\end{align}
Importantly, since the Hamiltonian is conserved, the same Hamiltonian generates time evolution for both the early- and late-time modes. 

It follows from \eqref{eq:hpsiflow} and \eqref{eq:hRscramblingflow} that if we write
\begin{equation}
    \beta h_R = \beta h_L - \log \Delta_\Psi\,,
\end{equation}
then we find that $[h_L, \A_R] = [h_L, \B_R] = 0$. The operator $h_L$ can again be identified with the left boundary Hamiltonian $H_L - \braket{H_L}$. The large $N$ algebra generated by $\A_R$, $\B_R$ and $h_R$ is just the crossed product
\begin{equation}
    \r^{\rm cr}\equiv \r\rtimes\R\,,
\end{equation}
which acts on the Hilbert space
\begin{equation}
    \H^{\rm cr} \cong\H_A\otimes\H_B\otimes L^2(\R)\,.
\end{equation}

Since the algebra $\r$ is a crossed product of the Type III$_1$ factor $\r$ by a modular flow, it is a Type II$_\infty$ factor and hence has a unique trace (up to rescaling). This trace can be written as
\begin{equation}\label{eq:trace}
    {\tr}[\mathcal{O}_R] = \int_{-\infty}^\infty dh_L e^{\beta h_L} \<\Psi|_A\<\Psi|_B \mathcal{O}_R |\Psi\>_A|\Psi\>_B\,,
\end{equation}
where $\mathcal{O}_R \in \r^{\rm cr}$ is regarded as a map from $h_L \in \R$ to operators on $\H_A \otimes \H_B$.

\subsection{Rényi 2‐entropies}
The trace \eqref{eq:trace} allows us to define density matrices $\rho_R$, by the condition that, for all $\mathcal{O}_R \in \r^{\rm cr}$,
\begin{align}
    {\tr}[\mathcal{O}_R \rho_R] = \braket{\mathcal{O}_R}\,.
\end{align}
We can then use those density matrices to define various entropies of states on $\r^{\rm cr}$. The most important of these entropies is the von Neumann entropy $S(\rho_R) = - \braket{\log \rho_R}$. However, this can often be difficult to work with in practice, and, for the algebra $\r^{\rm cr}$, it turns out to be very hard to compute for states other than the GNS vacuum. A  closely related, but sometimes computationally simpler, quantity is the Rényi 2‐entropy 
\begin{equation}\label{eq:S2}
    S_2(\rho_R) = -\log\left\<\rho_R\right\>\,.
\end{equation}

Even R\'{e}nyi 2-entropies are difficult to compute for generic states on $\r^{\rm cr}$. However, there is an important subclass of states for which we can make more progress, namely semiclassical states of the form
\begin{equation}
    |\widehat\Phi\> = \int dh_L \epsilon^{1/2} g(\epsilon h_L) |\Phi\>|h_L\>
\end{equation}
with $\ket{\Phi} \in \H_A \otimes \H_B$, $g \in L^2(\R)$ and $\varepsilon$ a finite but very small parameter. These states have $O(\varepsilon)$ fluctuations in the momentum conjugate to $h_L$, which describes the bulk timeshift between the left and right boundaries. In the bulk, they can therefore be described by matter fields in the state $\ket{\Phi}$ on an (almost) fixed classical background. 

The quantum extremal surface (QES) prescription says that the boundary von Neumann entropy of a semiclassical state is equal to the bulk generalized entropy, the sum of bulk entropy and area, of a region known as the entanglement wedge (bounded by the QES) \cite{Ryu:2006bv, Hubeny:2007xt, Wall:2012uf, Faulkner:2013ana, Engelhardt:2014gca, Headrick:2014cta}. In contrast, boundary R\'{e}nyi entropies  are typically dominated by a small tail of the wavefunction featuring a highly backreacted geometry. This tail of the wavefunction is difficult to treat semiclassically. 

In many ways, however, the more physically interesting, if smaller, contribution to R\'{e}nyi entropies is the part that comes from the peak of the wavefunction (i.e. the unbackreacted geometry). Like the von Neumann entropy, this contribution can be computed simply by replacing the density matrix $\rho_R$ in the formula \eqref{eq:S2} by the `generalized density matrix' associated to the entanglement wedge/QES \cite{Akers:2023fqr}. Moreover, when multiple extremal surfaces exist, the gravitational replica trick suggests that we should simply sum over contributions to the R\'{e}nyi 2-entropy of each QES. (In contrast, there is no known simple formula for the von Neumann entropy of states where there are multiple extremal surfaces with similar generalized entropy!)

The density matrix $\rho_\Phi$ of $\ket{\widehat\Phi}$ on $\r^{\rm cr}$ can be written, up to $\O(\epsilon)$ corrections, as \cite{Chandrasekaran:2022eqq}
\begin{equation}
    \rho_\Phi \approx \epsilon\bar g(\epsilon h_R)e^{-\beta h_L} \Delta_{\Phi|\Psi} g(\epsilon h_R) \,,
\end{equation}
where $\Delta_{\Phi|\Psi}$ is the relative modular operator of $\ket{\Phi}$ relative to $\ket{\Psi}$ on $\r$.\footnote{See \cite{Jensen:2023yxy} for the exact form of $\rho_\Phi$.} As expected, because of the factor of $e^{-\beta h_L}$, the R\'{e}nyi entropy of $\rho_{\Phi}$ is dominated by contributions from the exponentially small tail of the wavefunction with $h_L = -O(1/\epsilon)$. However, the contribution from the peak of the wavefunction, which we expect to have a semiclassical bulk interpretation, is given, up to $\O(\epsilon)$ corrections, by the formula \cite{Chandrasekaran:2022eqq}
\begin{equation}\label{eq:Renyiintegral}
    e^{-S_2(\rho_\Phi)} \approx \int_{-\infty}^\infty dh_L \epsilon^2|g(\epsilon h_L)|^4 e^{-\beta h_L} \<\Phi|\Delta_{\Phi|\Psi}|\Phi\> \,.
\end{equation}

For simplicity, as in \cite{Chandrasekaran:2022eqq}, we consider the two-shock states\footnote{The generalization to more than two shocks is straightforward but somewhat messy to write down explicitly.} of the form
\begin{equation}
    |\Phi\> = \sum_i a_{R,i}b_{R,i}|\Psi\>\,,
\end{equation}
with $a_{R,i} \in \A_R$ and $b_{R,i} \in \B_R$. The sum over $i$ allows the two shocks to be entangled with one another; from now on, we will use Einstein summation convention and implicitly sum over all repeated indices. The expectation value appearing in the integrand in \eqref{eq:Renyiintegral} can then be worked out explicitly as
\begin{align}
    \<\Phi|\Delta_{\Phi|\Psi} |\Phi\> &= \<\Phi|S_{\Phi|\Psi}^\dag S_{\Phi|\Psi}|\Phi\> = \<\Psi|b_{R,i}^\dag a_{R,i}^\dag S_{\Phi|\Psi}^\dag S_{\Phi|\Psi}a_{R,j}b_{R,j}|\Psi\> \nonumber\\
    &= \<\Psi|b_{R,i}^\dag a_{R,i}^\dag S_{\Phi|\Psi}^\dag b_{R,j}^\dag a_{R,j}^\dag|\Phi\> = \<\Phi|a_jb_j S_{\Phi|\Psi} a_{R,i}b_{R,i}|\Psi\> \nonumber\\
    &= \<\Phi|a_{R,j}b_{R,j} b_{R,i}^\dag a_{R,i}^\dag |\Phi\> = \<\Psi|b_{R,k}^\dag a_{R,k}^\dag a_{R,j}b_{R,j} b_{R,i}^\dag a_{R,i}^\dag a_{R,l}b_{R,l}|\Psi\>\,,
\end{align}
where, since the Tomita operator $S_{\Phi|\Psi}$ is antilinear, we have $\<x|S^\dag_{\Phi|\Psi} y\> = \overline{\<S_{\Phi|\Psi} x|y\>} = \<y| S_{\Phi|\Psi} x\>$. Finally, writing $a_{R,i} = e^{\ii P_A P_B/2} a_{i} e^{-\ii P_A P_B/2}$ and $b_{R,i} = e^{-\ii P_A P_B/2} b_{i} e^{\ii P_A P_B/2}$ explicitly, we obtain
\begin{align}\label{eq:Renyiformula}
    \<\Phi|\Delta_{\Phi|\Psi}|\Phi\> = \<\Psi|b_{k}^\dag a_{k}^\dag a_{j} e^{-\ii P_A P_B} b_{j} b_{i}^\dag e^{\ii P_A P_B} a_{i}^\dag a_{l} b_{l}|\Psi\> \,.
\end{align}

More generally, we can consider the state $\ket{\Psi_\alpha} = a_{R,i}b_{R,i}(t)|\Psi\>$ (with $\alpha = \exp(2\pi t)$ as in Section \ref{sec:limits}). We then have
\begin{align}\label{eq:Renyiformulaalpha}
    \<\Phi_\alpha|\Delta_{\Phi_\alpha|\Psi}|\Phi_\alpha\> = \<\Psi|b_{k}^\dag a_{k}^\dag a_{j} e^{-\ii \alpha P_A P_B} b_{j} b_{i}^\dag e^{\ii \alpha P_A P_B} a_{i}^\dag a_{l} b_{l}|\Psi\> \,.
\end{align}
In the tensor product limit of $\alpha\to0$, we recover the expected large $N$ factorization
\begin{equation}
    \lim_{\alpha\to0} \<\Phi_\alpha|\Delta_{\Phi_\alpha|\Psi}|\Phi_\alpha\> = \<a_{k}^\dag a_{l} a_{i}^\dag a_{j}\> \<b_{k}^\dag b_{l} b_{i}^\dag b_{j}\>\,.
\end{equation}
As explained in \cite{Chandrasekaran:2022eqq}, this formula matches the expected answer for the R\'{e}nyi entropy of a bulk state that contains a single QES, lying in the middle of both the left and right shocks. The corrections in $\alpha$ take the form
\begin{equation}
    \delta\<\Phi|\Delta_{\Phi|\Psi}|\Phi\> = \ii \alpha\<\Psi|b_{k}^\dag a_{k}^\dag a_{j} \left[b_{j} b_{i}^\dag, P_A P_B\right] a_{i}^\dag a_{l} b_{l}|\Psi\> + \O(\alpha^2) \,.
\end{equation}

In the free product limit of $\alpha\to\infty$, the Rényi 2‐entropy decomposes into a sum over contributions associated with different quantum extremal surfaces. To see this, we can use \eqref{eq:summutuallynoncrossing} to obtain
\begin{align}
    \lim_{\alpha\to\infty} \<\Phi_\alpha|\Delta_{\Phi_\alpha|\Psi}|\Phi_\alpha\> &= \< a_i^\dag a_j\> \< a_k^\dag a_l\> \< b_i^\dag b_jb_k^\dag b_l\> + \< a_i^\dag a_j a_k^\dag a_l\>_{\rm conn} \braket{b_jb_k^\dag} \braket{b_i^\dag b_l}\\
    &=  \< a_i^\dag a_j\> \< a_k^\dag a_l\> \< b_i^\dag b_jb_k^\dag b_l\> + \< b_i^\dag b_j\> \< b_k^\dag b_l\>  \left(\< a_i^\dag a_j a_k^\dag a_l\> - \< a_i^\dag a_j\> \< a_k^\dag a_l\>\right) \,. \label{eq:freeprodRenyi}
\end{align}
These three terms in \eqref{eq:freeprodRenyi} match the expected contributions from three distinct extremal surfaces, as was again explained in \cite{Chandrasekaran:2022eqq}. The two positive terms correspond to the expected contributions associated to the two ``throat'' surfaces, each of which lies in the middle of one shock and entirely in the past of the other shock. The magnitude of the negative term matches that expected from a ``bulge'' QES that lies almost entirely in the past of both shocks. The connection (if any) between the fact that this third QES is a bulge and the negative sign of its apparent contribution remains somewhat unclear.

It is interesting that the modular-twisted algebra $\r^{\rm cr}$ describes a smooth transition between these single- and multi-QES phases. We hope that a more careful analysis of \eqref{eq:Renyiformula} may provide some insight into how that transition plays out, but we leave that analysis to future work.

\section{Higher dimensions and localized excitations}\label{sec:local}
Up to this point, our construction has been confined, for simplicity, to two spacetime dimensions; in this section, we lift that restriction to study the algebraic structure of gravitational scrambling in arbitrary dimensions, including in particular localized excitations with nontrivial transverse profiles.\footnote{One might hope that in fact the original modular-twisted product algebra, without modification, also correctly describes the $s$-wave sector of higher-dimensional theories. Unfortunately this is only true for gravitational four-point functions, which involve a single gravitational $S$-matrix with only $s$-wave excitations in both the in- and out-states. In general, however, gravitational scattering only preserves the total angular momentum of the system and not the individual angular momenta of the ingoing and outgoing modes. As a result, modes of arbitrary angular momenta can be excited by scattering of $s$-wave modes and their effects can be seen in higher-point $s$-wave sector correlation functions.} We focus primarily on the case of an AdS-planar black hole, but the analysis can be extended beyond that fairly straightforwardly.

\subsection{Average null energies from boundary algebras}
Our starting point is the horizon algebra $\widetilde{\A}_{v(y)}$ associated to the future of a horizon cut $v(y)$.  Here $v(y)$ is the Eddington-Finkelstein infalling time of the cut as a function of the transverse coordinates $y$; we will also make frequent use below of the Kruskal coordinate $V = \exp(2 \pi v)$, which extends to $V < 0$ to describe the left white hole horizon. (For $v(y) = 0$, $\widetilde{\A}_{v(y)}$ is just the algebra $\widetilde \A$ defined previously.) The modular Hamiltonian $\widetilde K_A[v(y)] = - \log \widetilde{\Delta}_{A_{v(y)}}$ for this algebra is given by \cite{Wall:2011hj, Faulkner:2013ica, Jafferis:2015del}
\begin{equation}
    \widetilde K_A[v(y)] \equiv 2\pi \int_{-\infty}^\infty d^{d-2}y d V\,\, (V-V(y))\,T_{VV} \,.
\end{equation}
It follows immediately that the average null-energy operator is related to the functional derivative\footnote{This result has previously been used to probe the ANEC \cite{Ceyhan:2018zfg, Faulkner:2016mzt, Balakrishnan:2017bjg, Faulkner:2018faa} and the holographic dictionary \cite{Faulkner:2017vdd, Chandrasekaran:2021tkb}.}
\begin{equation}\label{eq:anefunctionalderiv}
    \frac{\delta \widetilde K_A}{\delta v(y)}\Big|_{v(y) = 0} = 2\pi \frac{\delta \widetilde K_A}{\delta V(y)}\Big|_{V(y) = 1} = - 4\pi^2 \int_{-\infty}^\infty d V\, T_{VV}\,.
\end{equation}

We can similarly define algebras $\widetilde{\A}_{t(y)}$ generated by asymptotic boundary operators to the future of a boundary cut $t(y)$. If we do so, we see that in fact $\widetilde{\A}_{v(y)}$ and $\widetilde{\A}_{t(y)}$ coincide not only when $t(y) = v(y) = 0$ (where they are both equal to $\widetilde\A$), but also at first order in $\delta t(y) \ll 1$ for $t(y) = v(y) = \delta t(y)$. To see this, it is sufficient to check that for any smooth function $t_0(y)$, the boundary of the future of the boundary cut $t(y) = \varepsilon\, t_0(y)$ intersects the horizon at $v(y) = \varepsilon\, t_0(y)$ up to a correction that is subleading in $\varepsilon$. It follows immediately that the functional derivatives of the modular Hamiltonians $\widetilde K_A[v(y)]$ (associated to $\widetilde{\A}_{v(y)}$) and $\widetilde K_A[t(y)]$ (associated to $\widetilde{\A}_{t(y)}$) agree at $v(y) = t(y) = 0$:
\begin{equation}\label{eq:Kv=Kt}
      \frac{\delta\widetilde K_A}{\delta t(y)} \Big|_{t(y)=0} =  \frac{\delta\widetilde K_A}{\delta v(y)} \Big|_{v(y)=0} =  - 4\pi ^2 \int_{-\infty}^\infty d V\, T_{VV}\,.
\end{equation}
Importantly, unlike $\widetilde{\A}_{v(y)}$, the algebra $\widetilde{\A}_{t(y)}$  has a simple boundary interpretation as the large $N$ von Neumann subalgebra generated by single-trace operators to the future of the cut $t(y)$. It follows that we can use \eqref{eq:Kv=Kt} to obtain a boundary interpretation of the average null-energy operator \eqref{eq:anefunctionalderiv}.

In exactly the same way, we can define an algebra $\widetilde\B_{u(y)}$ as the restriction of $\B$ to the past of the white hole horizon cut $u(y)$ and $\widetilde\B_{t(y)}$ as the restriction of $\B$ to the past of the boundary cut $t'(y)$ (i.e. $t(y) = t'(y) + T(N)$). Here $u(y)$ is defined so that $u(y) = {\rm const}$ is lightlike separated from the boundary cut with $t'(y) = u(y)$. For $u(y) = t'(y) = 0$, both of these algebras are the same as $\widetilde{\B}$. The functional derivatives of their modular Hamiltonians are
\begin{equation}\label{eq:Kv=Kt}
      \frac{\delta\widetilde K_B}{\delta t(y)} \Big|_{t(y)=0} = \frac{\delta\widetilde K_B}{\delta u(y)} \Big|_{u(y)=0} =  4\pi ^2 \int_{-\infty}^\infty d U\, T_{UU}\,,
\end{equation}
where the Kruskal coordinate $U$ satisfies $U = - \exp(-2 \pi u)$ for $U < 0$.

\subsection{The localized eikonal phase}
In $d$-dimensions the gravitational eikonal phase  is given by 
\begin{equation}
    \hat\delta = \frac{1}{N^2} e^{2\pi T} \int d^{d-2} x\,d^{d-2} y\,d U\,d V\, T_{UU}(U,x) T_{VV}(V,y) f(|x-y|) \,,
\end{equation}
where $f(|x-y|)$ is the transverse profile of the shock created by a high-energy localized source at $x$. This profile can be found by solving \cite{Roberts:2014isa, Shenker:2014cwa}
\begin{equation}
    (-\nabla^2+\mu^2)f(x) = \delta^{d-1}(x)\,,\qquad \mu^2 = \frac{2 \pi (d-1)}{\beta} \,,
\end{equation}
and for an AdS-planar black hole is given explicitly by
\begin{align}
    f(|x|) &= \frac{1}{2\pi} \left(\frac{\mu}{2\pi|x|}\right)^{\frac{d-3}{2}} K_{\frac{d-3}{2}}(\mu|x|) \,.
\end{align}
Cancelling the factors of $1/N^2$ and $\exp(2 \pi T)$ and absorbing $O(1)$ constants into the definition of $T$ where possible, we can write the eikonal phase as
\begin{equation} \label{eq:localizedeikonal}
    \hat\delta = -\int d x\,d y\,  \frac{\delta\widetilde K_A}{\delta t(x)} \Big|_{t(x)=0}~ \frac{\delta\widetilde K_B}{\delta t(y)} \Big|_{t(y)=0} ~f(|x-y|) \,.
\end{equation}

The gravitational scrambling algebra (generated as usual by $a_R \in \A_R$ and $b_R \in\B_R$) can then be defined, in close analogy to \eqref{eq:boostedelements}, to act on $\H_A \otimes \H_B$ by
\begin{equation} 
\begin{split}
    a_R &= e^{\ii \hat\delta/2} \,a\,e^{-\ii \hat\delta/2} \,,\\
    b_R &= e^{-\ii \hat\delta/2}\,b\,e^{\ii \hat\delta/2} \,, 
\end{split}
\end{equation} 
with the operator $\hat\delta$ given by \eqref{eq:localizedeikonal}, while $a \in \A $ and
$b \in \B$.

We expect that replacing the phase $\hat\delta = P_A P_B$ appearing in the original modular-twisted product algebra by \eqref{eq:localizedeikonal} will not change the algebraic properties that we proved in Sections \ref{sec:main} and \ref{sec:limits}. When integrated against any positive smearing function, the operators  $-\delta\widetilde K_{A}/\delta t(x)$ and $\delta\widetilde K_{B}/\delta t(x)$ become the generators of half-sided modular translations. By expanding the integration kernel $f$ as a sum of products of positive functions, the phase \eqref{eq:localizedeikonal} can therefore be approximated to arbitrary precision by a (finite) sum over products of half-sided modular translation generators. We are not aware of any significant issues that arise for the proofs in Sections \ref{sec:main} and \ref{sec:limits} if the product $P_A P_B$ is replaced by a sum over such products.

\subsection*{Acknowledgements}
We especially thank Chang-Han Chen for initial collaboration and many useful discussions. We would also like to thank Gabriele Di Ubaldo, Felix Haehl, Steve Shenker, Jon Sorce, Douglas Stanford, Zhenbin Yang, and Shunyu Yao for valuable discussions. GP is supported by the Department of Energy through DE-SC0019380 and DE-FOA-0002563, by AFOSR award FA9550-22-1-0098 and by a Sloan Fellowship.

\bibliographystyle{JHEP}
\bibliography{ref}
\end{document}